\documentclass[twocolumn,epjc3]{svjour3}

\RequirePackage[T1]{fontenc}
\RequirePackage{mathptmx}      
\RequirePackage{flushend}
\RequirePackage[numbers,sort&compress]{natbib}
\RequirePackage[colorlinks,citecolor=blue,urlcolor=blue,linkcolor=blue]{hyperref}

\journalname{Eur. Phys. J. C}

\usepackage{amsmath}
\usepackage{amssymb}
\usepackage{booktabs}
\usepackage{graphicx}
\usepackage{listings}
\usepackage{multirow}
\usepackage{xcolor}

\usepackage{textcomp}

\allowdisplaybreaks[1]

\numberwithin{equation}{section}
\def \refeq#1{(\ref{#1})}
\def \refsec#1{Section~\ref{#1}}
\def \refapp#1{\ref{#1}}
\def \reffig#1{Figure~\ref{#1}}
\def \reftab#1{Table~\ref{#1}}

\def \nn{\nonumber\\}

\newcommand{\checked}[1]{{\color{brown}{ {\bf Checked: }{#1}}}}
\renewcommand{\checked}[1]{#1}

\newcommand{\wilson}[2][{}]{\mathcal{C}_{#2}^{\mathrm{#1}}}

\renewcommand{\[}{\big[}
\renewcommand{\]}{\big]}

\renewcommand{\)}{\big)}
\newcommand{\dd}{{\rm d}}

\newcommand{\GeV}{\ensuremath{\mathrm{GeV}}}

\def \vecth{\vec{\theta}}

\def \cLdB1{{{\cal L}_{\Delta B = 1}^{\rm EW}}} 
\def \BR{{\cal B}}                               

\def \Op{{\cal O}}

\def \One{\leavevmode\hbox{\small1\kern-3.6pt\normalsize1}} 

\newcommand\rmdx[1]{\mbox{d} \, #1 \,}

\def \alE{\alpha_e}        

\def \B0toK0ast{{ \bar{B}^0 \to K^{\ast 0}}}
\def \B0toK0mumu{{ \bar{B}^0 \to K^0 \bar{\mu} \mu}}
\def \barB0toKKpill{{ \bar{B}^0 \to \bar{K}^{\ast 0} (\to K^- \pi^+) \bar{l}l}}
\def \B0toKKpill{{ B^0 \to K^{\ast 0} (\to K^+ \pi^-) \bar{l}l}}

\newlength{\relwidth}

\def \eos{\texttt{EOS}}
\begin{document}

\sloppy

\title{Constraints on tensor and scalar couplings from
  $B\to K\bar\mu\mu$ and $B_s\to \bar\mu\mu$
}
\author{Frederik Beaujean\thanksref{e1,addr1}
  \and
  Christoph Bobeth\thanksref{e2,addr2}
  \and
  Stephan Jahn\thanksref{e3,addr3,addr4}
}
\institute{
  C2PAP, Excellence Cluster Universe,
  Ludwig-Maximilians-Universit\"at M\"unchen,
  Garching, Germany\label{addr1}
  \and
  Institute for Advanced Study,
  Technische Universit\"at M\"unchen,
  Garching, Germany\label{addr2}
  \and
  Excellence Cluster Universe,
  Technische Universit\"at M\"unchen,
  Garching, Germany\label{addr3}
  \and
  \emph{Present Address:}
  Max Planck institute for physics, Munich, Germany\label{addr4}
}

\thankstext[$\star$]{t1}{{Preprint: EOS-2015-03,} FLAVOUR(267104)-ERC-107}
\thankstext{e1}{e-mail: frederik.beaujean@lmu.de}
\thankstext{e2}{e-mail: christoph.bobeth@ph.tum.de}
\thankstext{e3}{e-mail: sjahn@mpp.mpg.de}

\maketitle

\begin{abstract}
  The angular distribution of $B\to K\bar\ell\ell$ ($\ell =
  e,\,\mu,\,\tau$) depends on two parameters, the lepton
  forward-backward asymmetry, $A_{\rm FB}^\ell$, and the flat term,
  $F_H^\ell$. Both are strongly suppressed in the standard model and
  constitute sensitive probes of tensor and scalar contributions. We
  use the latest experimental results for $\ell = \mu$ in combination
  with the branching ratio of $B_s\to \bar\mu\mu$ to derive the
  strongest model-independent bounds on tensor and scalar effective
  couplings to date. The measurement of $F_H^\mu$ provides a
  complementary constraint to that of the branching ratio of $B_s\to
  \bar\mu\mu$ and allows us---for the first time---to constrain all
  complex-valued (pseudo-)scalar couplings and their chirality-flipped
  counterparts in one fit. Based on Bayesian fits of various
  scenarios, we find that our bounds even become tighter when vector
  couplings are allowed to deviate from the standard model and that
  specific combinations of angular observables in $B \to K^*$ are
  still allowed to be up to two orders of magnitude larger than in the
  standard model, which would place them in the region of LHCb's
  sensitivity.
\end{abstract}

%
%
%
\section{
 \checked{ Introduction}
}

With the analysis of the data collected by the LHCb Collaboration
during run I at the Large Hadron Collider (LHC), we now have access to
rather large samples of rare $B$-meson decays with branching ratios
below $10^{-5}$. As a consequence, angular analyses of three- and
four-body final states can be used to measure a larger number of
observables than previously possible at the B factories BaBar and
Belle. In this work we focus on rare $B$ decays driven at the parton
level by the flavor-changing neutral-current (FCNC) transition $b\to s
\bar\ell\ell$ that constitutes a valuable probe of the standard model
(SM) and provides constraints on its extensions.

The angular distribution of $B\to K \bar\ell\ell$---normalized to the
width $\Gamma_\ell$---in the angle $\theta_\ell$ between $B$ and
$\ell^-$ as measured in the dilepton rest frame is
\begin{align}
  \label{eq:ang-distr}
  \frac{1}{\Gamma_\ell} \frac{\mbox{d}\Gamma_{\ell}}{\mbox{d}\!\cos\theta_\ell} &
  = \frac{3}{4} (1 - F_H^\ell) \sin^2\!\theta_\ell
  + \frac{1}{2} F_H^\ell
  + A_{\rm FB}^\ell \cos\theta_\ell \,.
\end{align}
LHCb analyzed their full run 1 data set of 3 fb$^{-1}$ and measured the angular
distribution of the mode $B^+\to K^+ \bar\mu\mu$, i.e. $\ell =
\mu$~\cite{Aaij:2014tfa}, with unprecedented precision. They provide the
lepton-forward-backward asymmetry $A_{\rm FB}^\mu$ and the flat term $F_H^\mu$
in CP-averaged form and integrated over several bins in the dilepton invariant
mass $q^2$. Similarly, the CP-averaged branching ratios, ${\cal B}_\mu =
\tau_{B} \Gamma_\mu$,~\cite{Aaij:2014pli} and the rate CP asymmetry $A_{\rm
  CP}^\mu$ \cite{Aaij:2014bsa} are also available from 3 fb$^{-1}$.

Both {angular} observables, $F_H^\ell$ and $A_{\rm FB}^\ell$, exhibit strong suppression
factors for vector and dipole couplings present in the SM, thereby enhancing
their sensitivity to tensor and scalar couplings \cite{Bobeth:2007dw,
  Bobeth:2012vn}. A similar enhancement of scalar couplings compared to
helicity-suppressed vector couplings of the SM is well-known from $B_s\to
\bar\mu\mu$. Unfortunately the limited data set of $B\to K^*(\to K\pi)
\bar\ell\ell$ from LHCb \cite{Aaij:2013qta} {did not yet} allow to perform a full
angular analysis without the assumption of vanishing scalar and tensor couplings
in this decay mode. In the future with more data or special-purpose analysis
techniques like the method of moments~\cite{beaujean:2015mom}, certain angular
observables in $B\to K^* \bar\ell\ell$ will provide additional constraints on
such couplings, as for example $J_{6c}$ \cite{Altmannshofer:2008dz} and the
linear combinations $(J_{1s} - 3 J_{2s})$ and $(J_{1c} + J_{2c})$
\cite{Matias:2012xw, Bobeth:2012vn} as well as the experimental test of the
relations $H_T^{(2)} = H_T^{(3)}$ and $J_7 = 0$~\cite{Bobeth:2012vn} at low
hadronic recoil.

Here we exploit current data from $B^+\to K^+ \bar\mu\mu$ and $B_s\to
\bar\mu\mu$ to derive stronger constraints than before on tensor and scalar
couplings in various model-independent scenarios and study their impact on the
not-yet-measured sensitive observables in $B\to K^* \bar\ell\ell$. In
\refsec{sec:EFT}, we specify the effective theory of $|\Delta B| = 1$ decays on
which our model-independent fits are based. Within this theory, we discuss the
dependence of observables in $B\to K \bar\ell\ell$ and $B\to K^* \bar\ell\ell$
on the tensor and scalar couplings in \refsec{sec:observables} and specify also
the experimental input used in the fits. The constraints on tensor and scalar
couplings from the data are presented for several model-independent scenarios in
\refsec{sec:results}.  Technical details of the angular observables in $B\to K^*
\bar\ell\ell$, the branching fraction of $B_s\to \bar\mu\mu$, the treatment of
theory uncertainties, and the Monte Carlo methods used are relegated to
appendices.

%
%
%
\section{
  \checked{Effective Theory}
  \label{sec:EFT}
}

In the framework of the $|\Delta B| = |\Delta S| = 1$ effective theory
\begin{equation}
  \label{eq:Heff}
\begin{aligned}
  {\cal L}_{\rm eff}  =
   \frac{4 G_F}{\sqrt{2}} \,\frac{\alE}{4 \pi}\, & V_{tb}^{} V_{ts}^\ast
      \, \Big[ {C_7(\mu_b) \Op_7 + C_{7'}(\mu_b) \Op_{7'}}
 \\ & +
       \sum_{\ell = e,\, \mu,\, \tau} \sum_i  \wilson[\ell]{i}(\mu_b)
       {\Op_i^\ell} \Big] + \text{h.c.}
   {\,,}
\end{aligned}
\end{equation}
the most general dimension-six flavor-changing operators {$\Op_i^{(\ell)}$} mediating $b\to
s \gamma$ and $b\to s \bar\ell\ell$ are classified according to their chiral
structure. There are dipole ($i = 7,7'$) and vector ($i = 9,9',10,10'$) operators
\begin{equation}
  \label{eq:SM:ops}
\begin{aligned}
  \Op_{7(7')} &
  = \frac{m_b}{e}\!\[\bar{s} \sigma^{\mu\nu} P_{R(L)} b\] F_{\mu\nu}\,,
\\
  {\Op_{9(9')}^\ell} &
  = \[\bar{s} \gamma_\mu P_{L(R)} b\]\!\[\bar{\ell} \gamma^\mu \ell\]\,,
\\[0.1cm]
  {\Op_{10(10')}^\ell} &
  = \[\bar{s} \gamma_\mu P_{L(R)} b\]\!\[\bar{\ell} \gamma^\mu \gamma_5 \ell\]\,,
\end{aligned}
\end{equation}
further scalar ($i = S,S',P,P'$) operators
\begin{equation}
  \label{eq:scalar:ops}
\begin{aligned}
  {\Op_{S(S')}^\ell} &
  = \[\bar{s} P_{R(L)} b\]\!\[\bar{\ell} \ell\]\,,
\\[0.1cm]
  {\Op_{P(P')}^\ell} &
  = \[\bar{s} P_{R(L)} b\]\!\[\bar{\ell} \gamma_5 \ell\]\,,
\end{aligned}
\end{equation}
and tensor ($i = T,T5$) operators
\begin{equation}
  \label{eq:tensor:ops}
\begin{aligned}
  {\Op_{T}^\ell} &
  = \[\bar{s} \sigma_{\mu\nu} b\]\!\[\bar{\ell} \sigma^{\mu\nu} \ell\]\,,
\\[0.1cm]
  {\Op_{T5}^\ell} &
  = \[\bar{s} \sigma_{\mu\nu} b\]\!\[\bar{\ell} \sigma^{\mu\nu} \gamma_5 \ell\]\,,
\end{aligned}
\end{equation}
where the notation ${\Op_{T5}^\ell} = i/2\, \varepsilon^{\mu\nu\alpha\beta}
\[\bar{s} \sigma_{\mu\nu} b\]\!\[\bar{\ell} \sigma_{\alpha\beta} \ell\]$ is
also used frequently in the literature. The respective short-distance couplings,
the Wilson coefficients $\wilson[(\ell)]{i}(\mu_b)$ are evaluated at a scale of
the order of the $b$-quark mass $\mu_b \sim m_b$ and can be modified from SM
predictions in the presence of new physics.

The SM values $\wilson[(\ell)]{7,9,10}$ are obtained at
next-to-next-to leading order (NNLO) \cite{Bobeth:2003at,
  Huber:2005ig} and depend on the fundamental parameters of the
top-quark and $W$-boson masses, as well as on the sine of the weak
mixing angle.  Moreover, they are universal for the three lepton
flavors $\ell = e,\, \mu,\, \tau$.  All other Wilson coefficients are
numerically suppressed or zero: $\wilson[SM]{7'} = m_s/m_b\,
\wilson[SM]{7}$, $\wilson[\ell, SM]{S,S',P,P'} \sim {m_b
  m_\ell/m_W^2}$, and $\wilson[\ell, SM]{9',10',T,T5} = 0$.  The
Wilson coefficients of the four-quark current-current and QCD-penguin
operators as well as of the chromomagnetic dipole operators are set to
their NNLO SM values at $\mu_b = 4.2 \, \mbox{GeV}$
\cite{Bobeth:2003at, Huber:2005ig}.

For the rest of this article, we will suppress the lepton-flavor index
on the Wilson coefficients $\wilson[\ell]{i} \to \wilson[]{i}$ and
operators $\Op_i^\ell \to \Op_i$. In \refsec{sec:results} we exploit
data with $\ell =\mu$ only, hence all derived constraints apply in
principle only to the muonic case but can be carried over to the other
lepton flavors $\ell = e,\, \tau$ for NP models that do not violate
lepton flavor. In general, the Wilson coefficients are decomposed into
SM and NP contributions $\wilson[]{i} = \wilson[SM]{i} +
\wilson[NP]{i}$ but often we will use $\wilson[]{i}$ for Wilson
coefficients with zero (or suppressed) SM contributions synonymously
with $\wilson[NP]{i}$.

%
%
%
\section{
  \checked{Observables and experimental input}
  \label{sec:observables}
}

The full dependence of $F_H^\ell$ and $A_{\rm FB}^\ell$ on tensor and
scalar couplings has been presented in \cite{Bobeth:2007dw,
  Bobeth:2012vn}, adopting the effective theory \refeq{eq:Heff},
i.e. neglecting higher-dimensional operators with {$\mbox{dim} \geq
  8$}. These results imply that for SM values of the effective
couplings
\begin{align}
  F_H^\ell(q^2) \big|_{\rm SM} & \propto \frac{m_\ell^2}{q^2}
    \bigg/ \Gamma_\ell(q^2) \,, &
  A_{\rm FB}^\ell(q^2) \big|_{\rm SM} & = 0 \,.
\end{align}
Hence, for $\ell = e,\, \mu$ both observables are quasi-null tests.
The flat term $F_H^\ell(q^2)|_{\rm SM}$ is strongly suppressed {by
  small lepton masses for the considered kinematic region $1 \leq
  q^2 \leq 22$ GeV$^2$} \cite{Bobeth:2007dw, Bobeth:2011nj,
  Bouchard:2013eph}.  Nonzero values of $A_{\rm FB}^\ell|_{\rm SM}$
can be induced by higher-order QED corrections, which will modify the
simple $\cos\theta_\ell$ dependence of the angular
distribution~\refeq{eq:ang-distr}, however, currently there is no
solid estimate available for this source of SM background\footnote{
  Logarithmically enhanced NLO QED corrections to $B\to X_s
  \bar\ell\ell$~\cite{Huber:2015sra} turn out to be non-negligible
  for angular observables,
  however, analogous corrections to $B\to K \bar\ell\ell$ are partially
  included in the event simulation of the experimental
  analysis~\cite{Aaij:2014tfa} via PHOTOS~\cite{Golonka:2005pn}.
  As proposed in \cite{Gratrex:2015hna}, the measurement of higher
  moments of the decay distribution~\refeq{eq:ang-distr} could give an
  estimate of the size of higher-order QED corrections, but still
  admixed with contributions of $\mbox{dim} \geq 8$ operators.}.
This picture does not change in the presence of new physics contributions to
vector and dipole operators $i = 7,7',9,9',10,10'$. On the other hand,
nonvanishing tensor or scalar contributions are enhanced unless the dynamics of
NP implies similar suppression factors, i.e., lepton-Yukawa couplings for
$F_H^\ell$ or $\alpha_e$ in the case of $A_{\rm FB}^\ell$. In particular,
$F_H^\ell$ is very sensitive to tensor couplings (see \refeq{eq:AFB-dependence}
below) and $A_{\rm FB}^\ell$ is sensitive to the interference of tensor and
scalar couplings (see \refeq{eq:FH-dependence} below).

\begin{figure*}
  \setlength{\relwidth}{0.33\textwidth}
  \begin{center}
    \includegraphics[width=\relwidth]{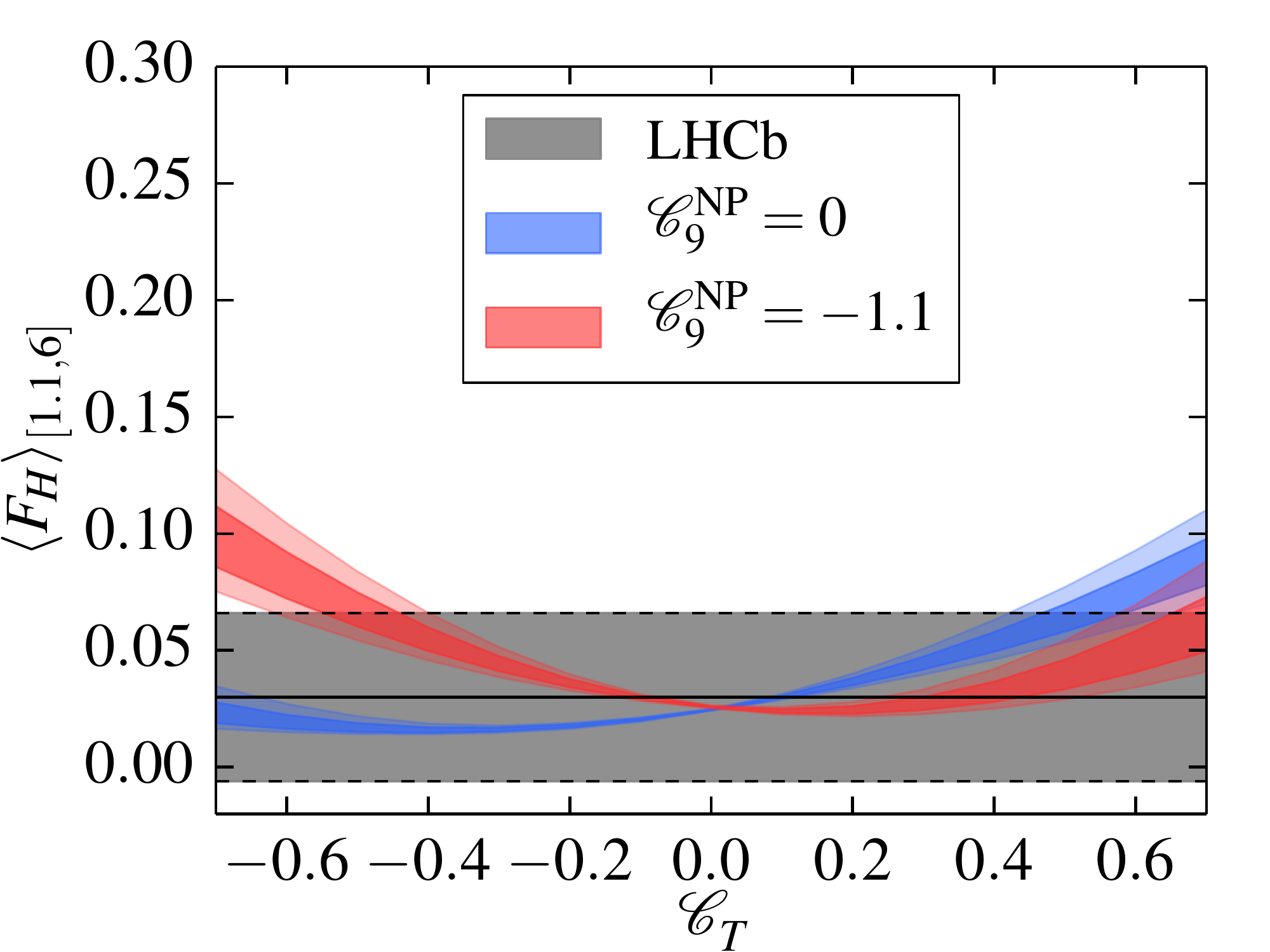}
    \includegraphics[width=\relwidth]{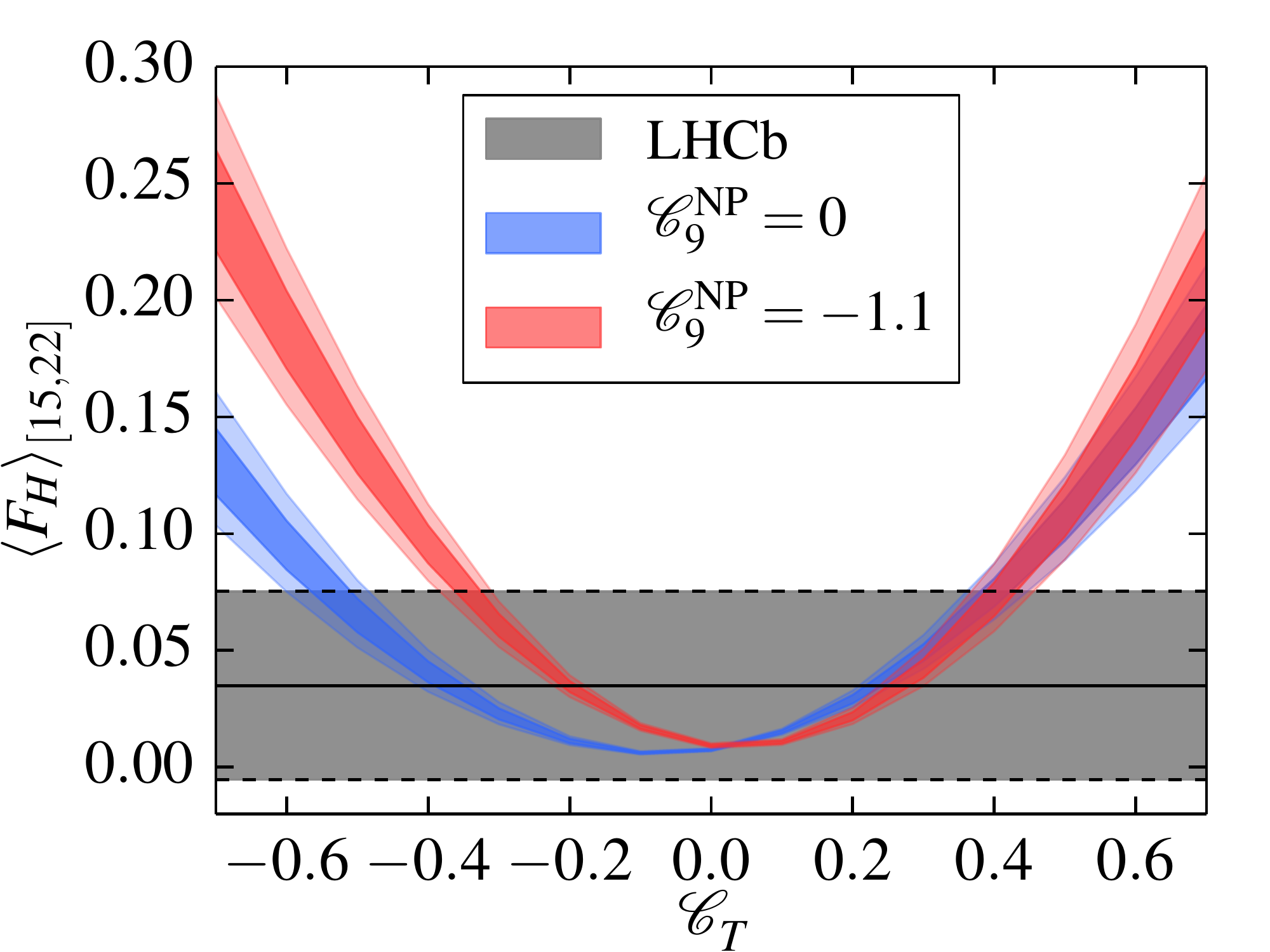}
    \\
    \includegraphics[width=\relwidth]{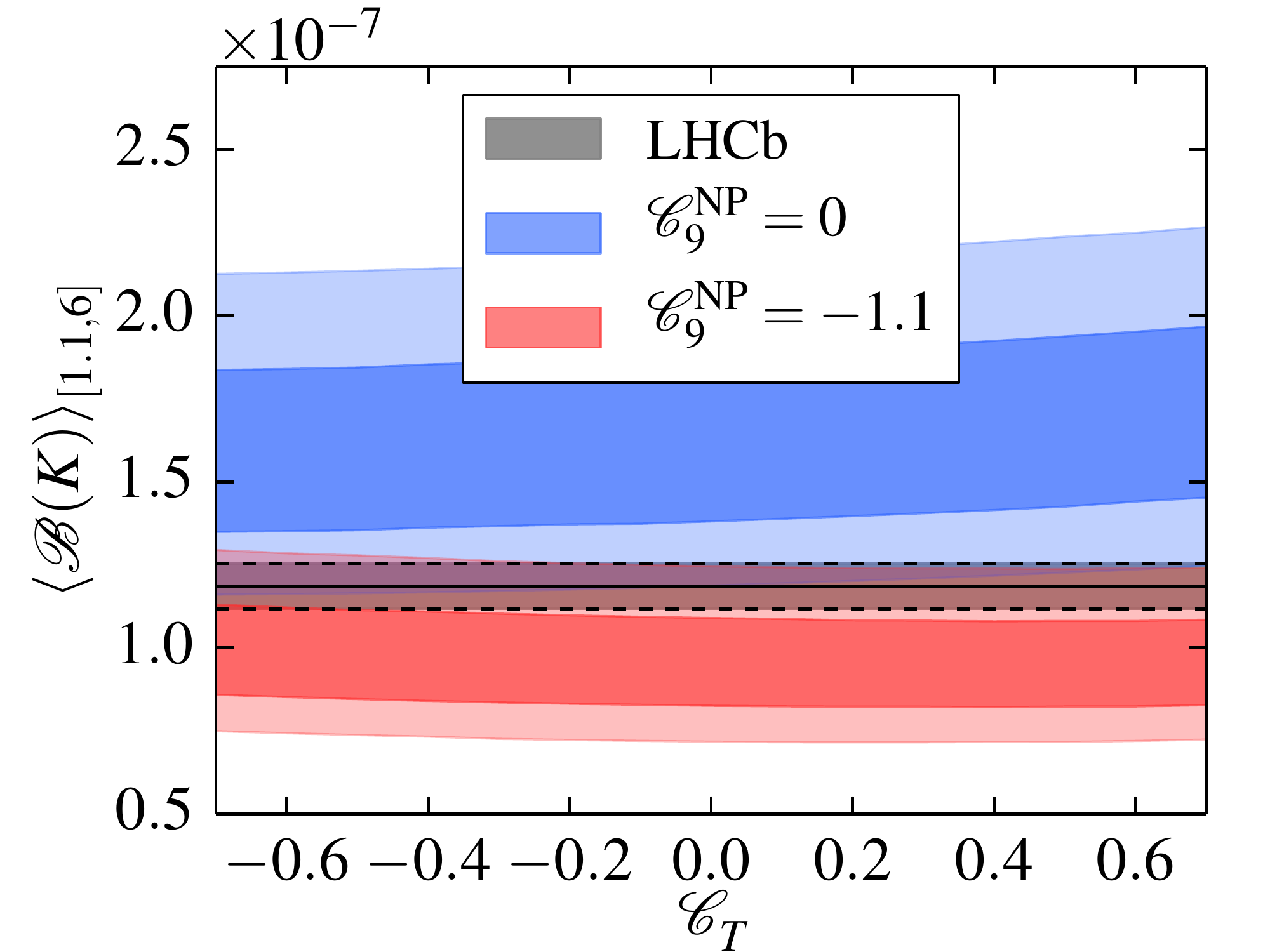}
    \includegraphics[width=\relwidth]{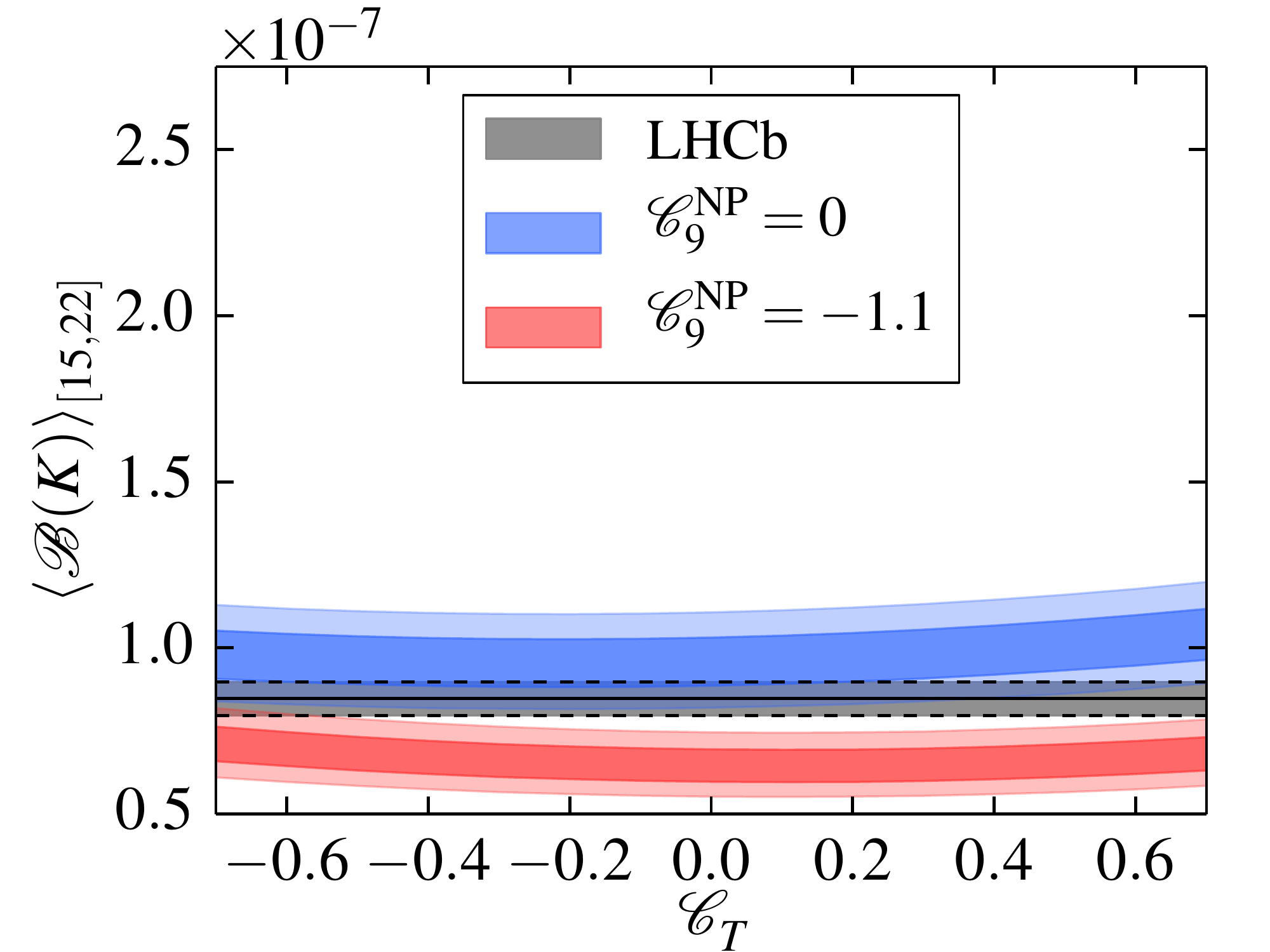}
    \\
    \includegraphics[width=\relwidth]{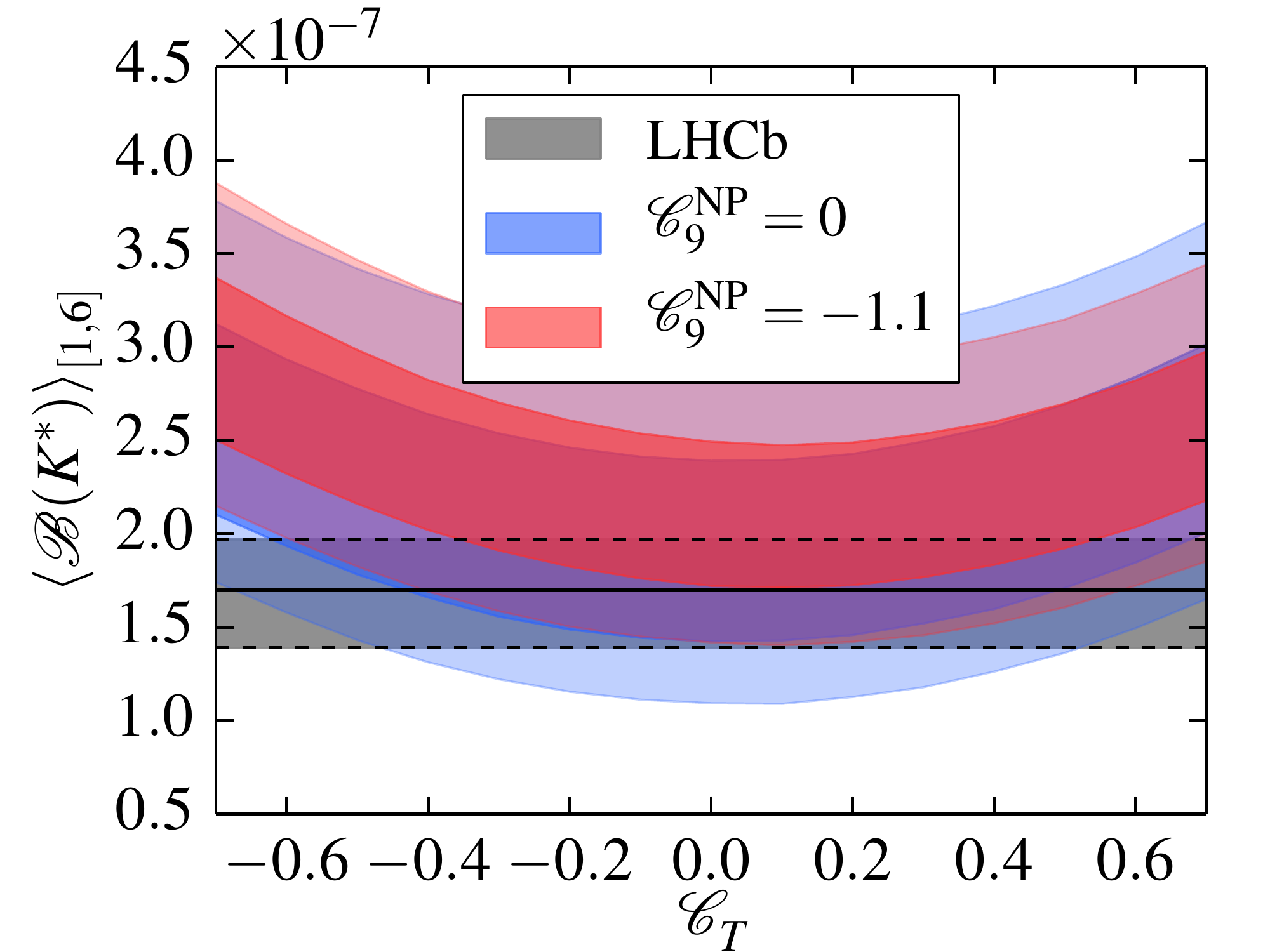}
    \includegraphics[width=\relwidth]{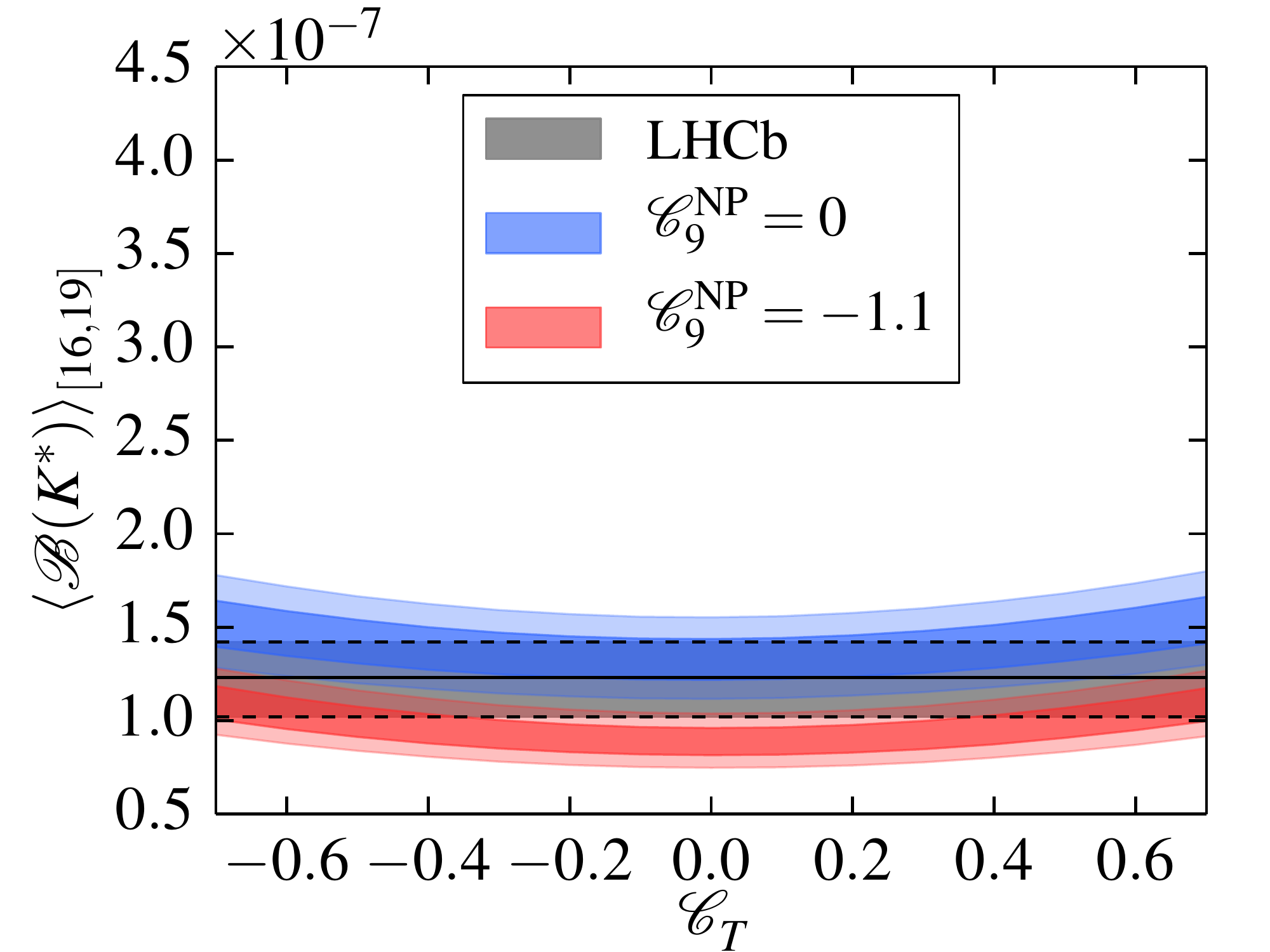}
    \\
    \includegraphics[width=\relwidth]{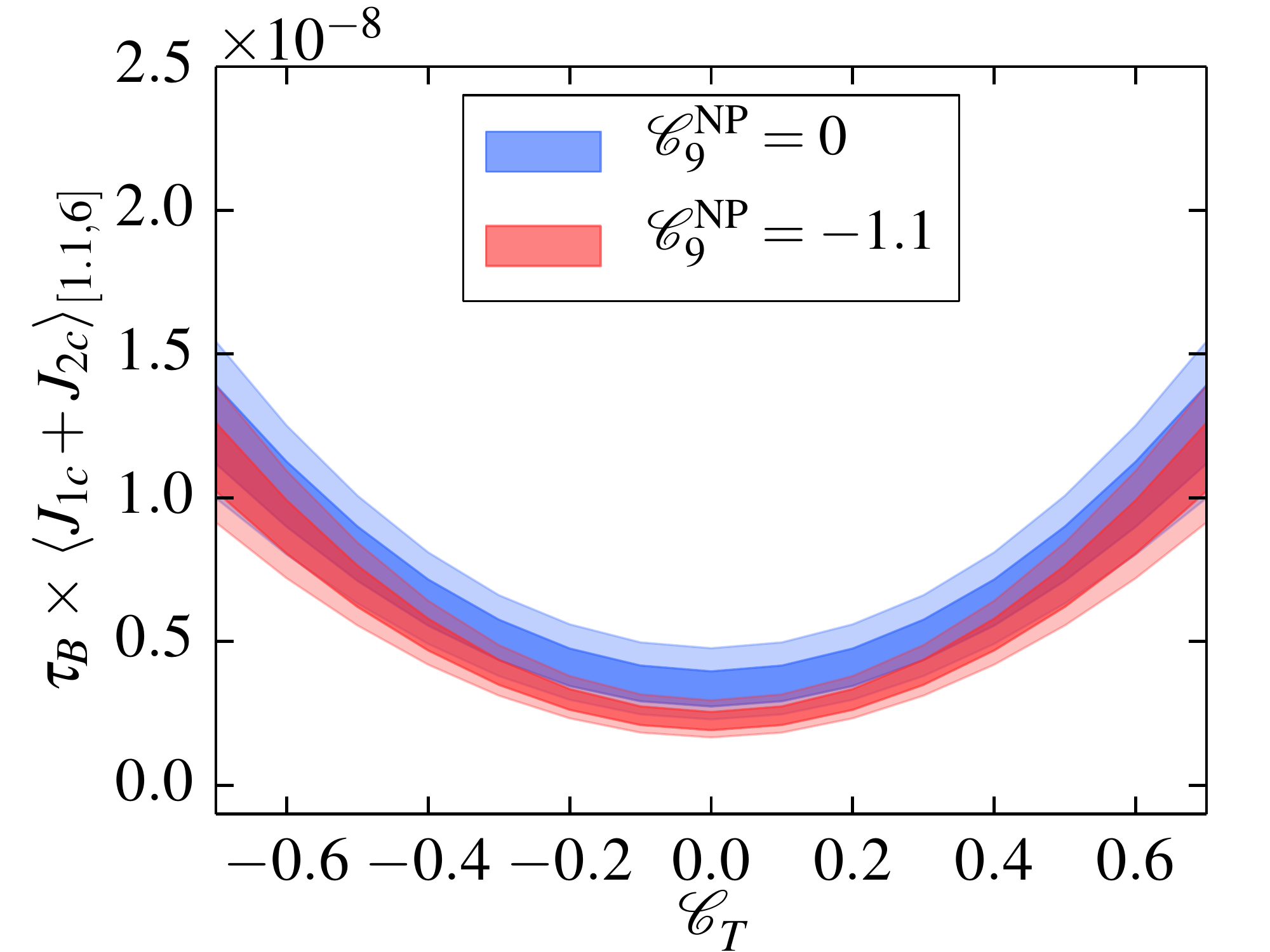}
    \includegraphics[width=\relwidth]{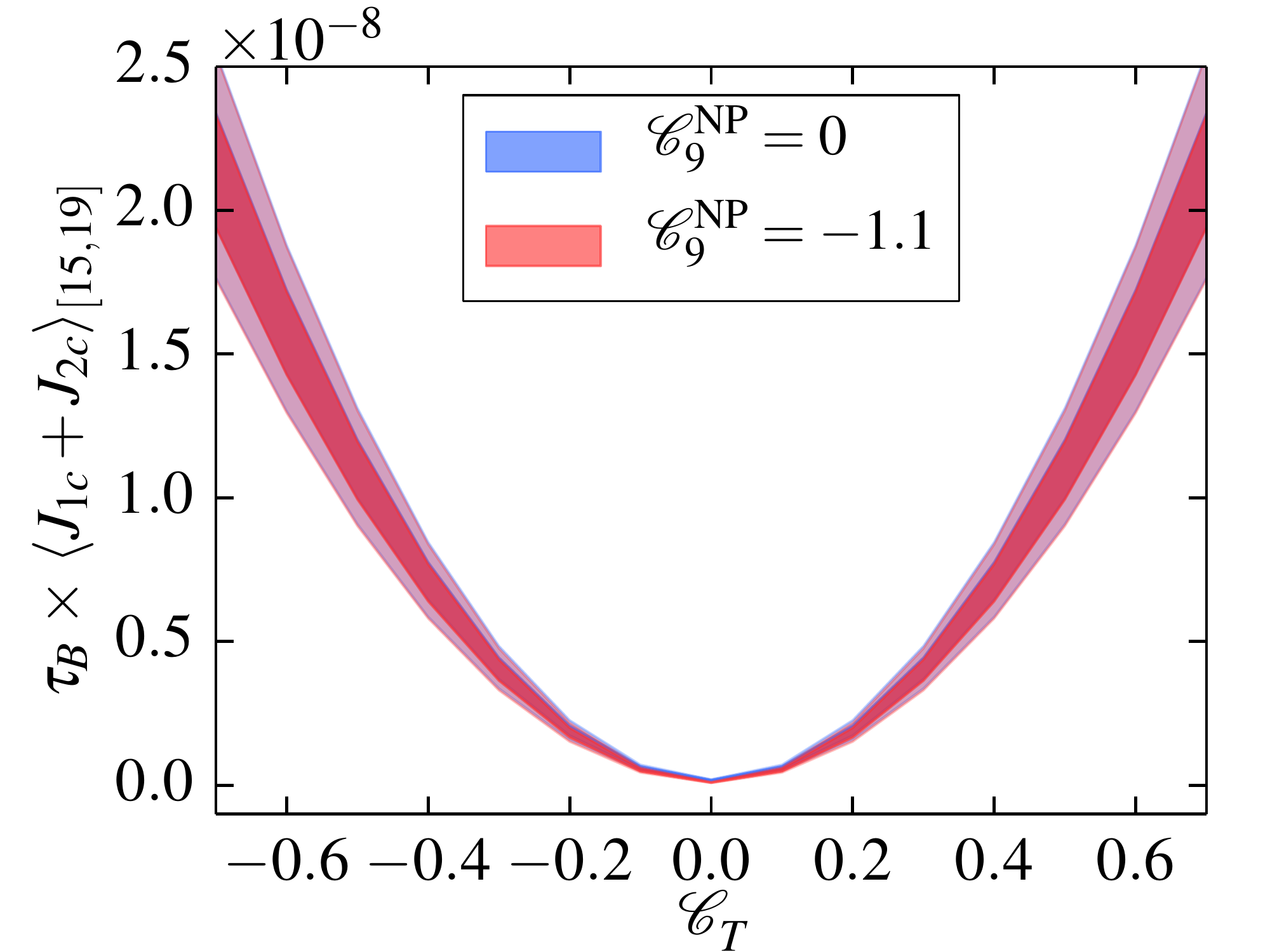}
    \\
    \includegraphics[width=\relwidth]{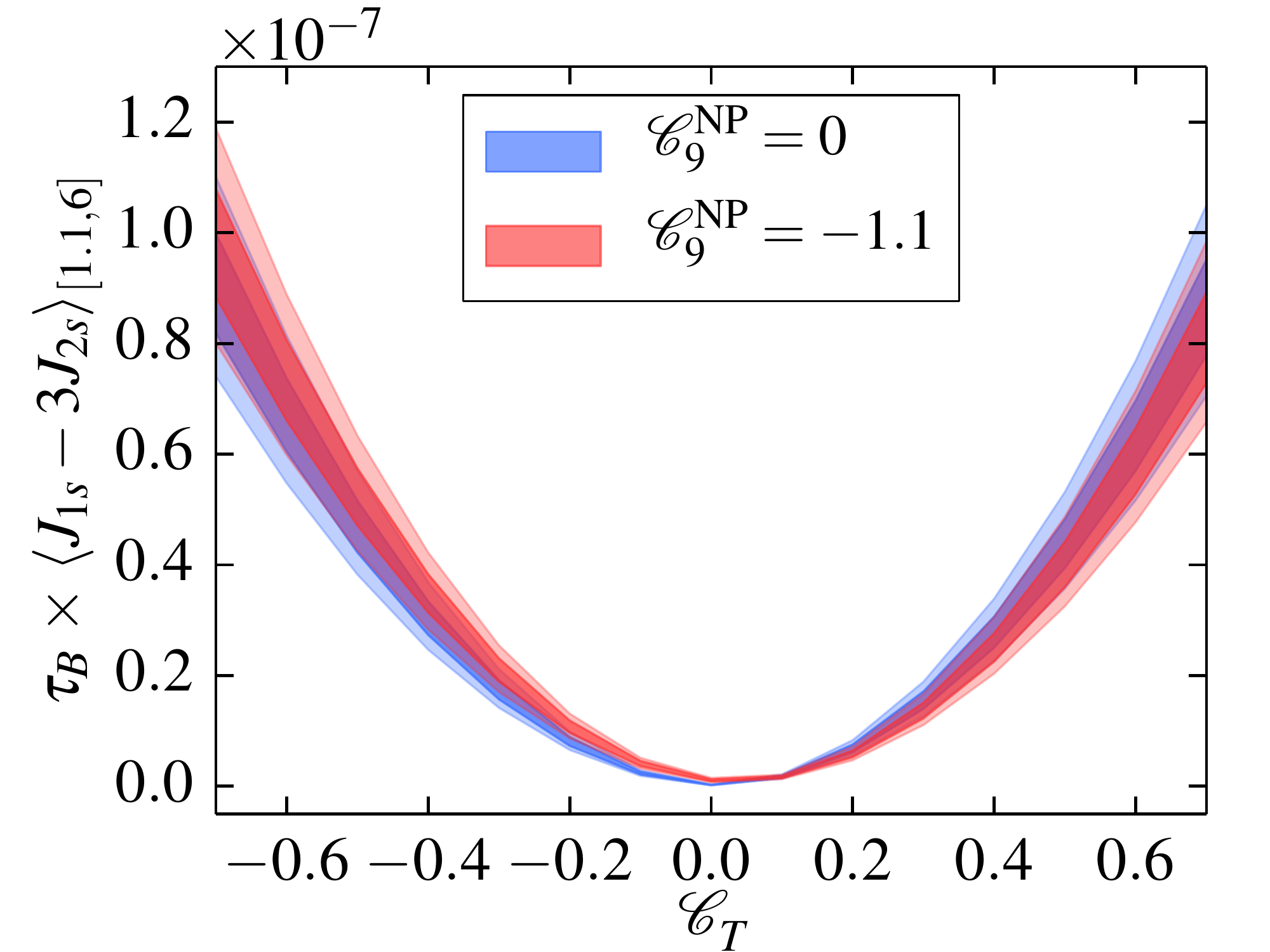}
    \includegraphics[width=\relwidth]{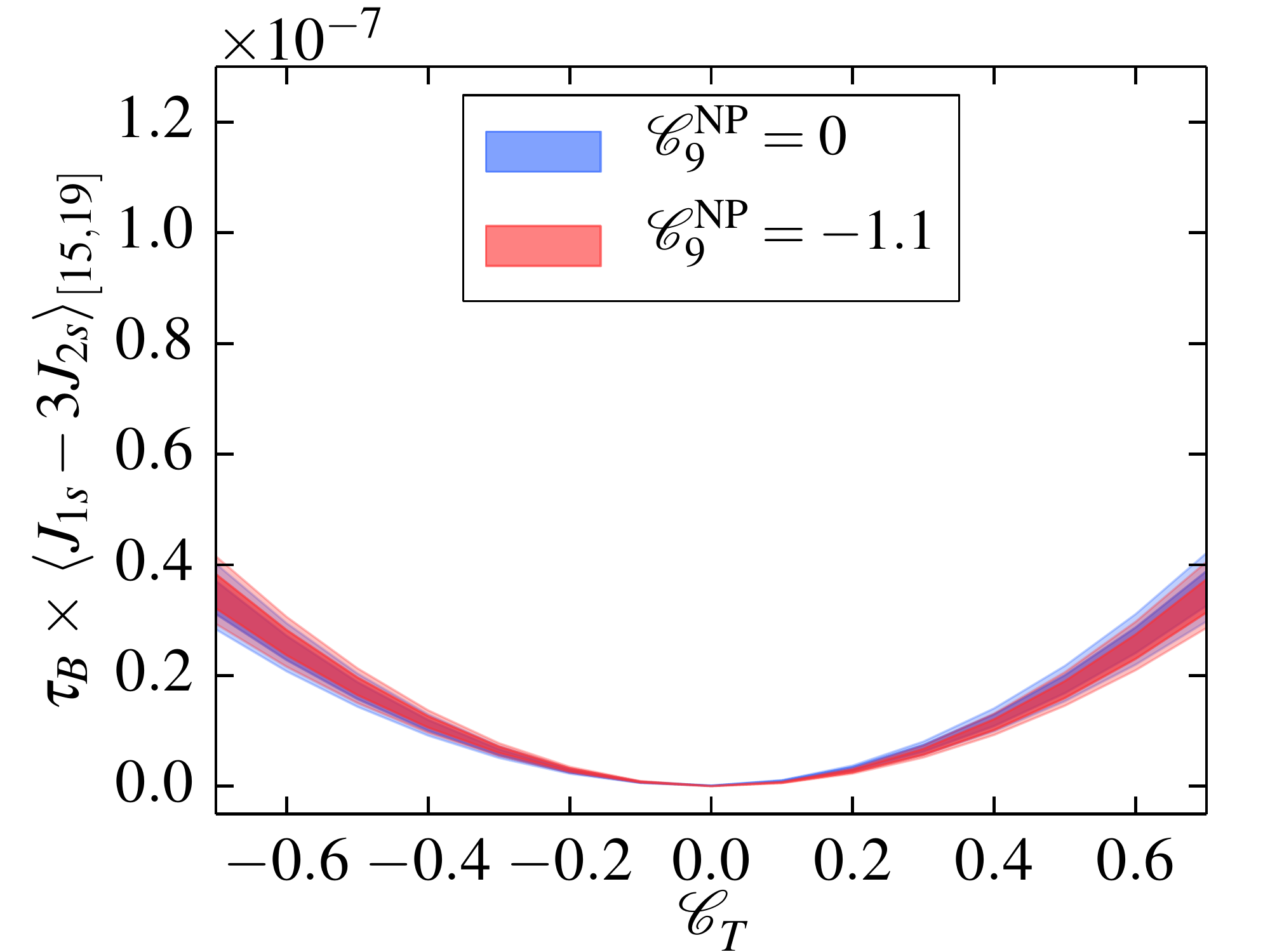}
  \end{center}
  \caption{ The sensitivity to the tensor coupling
    $\mbox{Re}(\wilson{T})$ of $F_H^\mu$ and {${\cal B}_\mu$} in $B^+
    \to K^+ \bar\mu\mu$ {as well as} {${\cal B}_\mu$}, $(J_{1c} +
    J_{2c})$, and $(J_{1s}-3 J_{2s})$ in $B^0 \to K^{*0}
    \bar\mu\mu$. Angular observables are rescaled by the lifetime of
    the $B$ meson, $\tau_B$. The bands represent the theory
    uncertainties at 68\% and 95\% probability of the prior
    predictive.  Two sets of bands are shown for $\wilson[NP]{9} = 0$
    (blue) and $\wilson[NP]{9} = -1.1$ (red). If available, the gray
    band indicates the latest 68\% confidence interval reported by
    LHCb. All observables are integrated over $q^2 \in
    [q_1^2,\,q_2^2]$ bins denoted as $\langle \dots \rangle_{[q^2_1,\,
      q^2_2]}$ to match LHCb. }
  \label{fig:FHvsCT}
\end{figure*}

There are some angular observables $J_i$ in $B\to K^*(\to K\pi)\, \bar\ell\ell$
with the same properties; i.e. tensor and scalar contributions are kinematically
enhanced by a factor $\sqrt{q^2}/m_\ell$ over vector ones present in the SM or
their respective interference terms. These are $J_{6c}$ and the two linear
combinations $(J_{1s} - 3 J_{2s})$ and $(J_{1c} + J_{2c})$ with explicit
formulas given in \refapp{app:BKstarellell}. In our fits and predictions we
include all kinematically suppressed terms. But for the purpose of illustration,
we now consider the analytical dependence for vanishing lepton mass. In this
limit,
\begin{align}
  J_{6c} & \propto
    \mbox{Re}\big[(\wilson[]{P} - \wilson[]{P'})\, \wilson[\ast]{T}
                - (\wilson[]{S} - \wilson[]{S'})\, \wilson[\ast]{T5}\big]
\intertext{is sensitive to the interference of tensor and scalar operators,
  complementary to $A_{\rm FB}^\ell$ in $B\to K \bar\ell\ell$}
  \label{eq:AFB-dependence}
  A_{\rm FB}^\ell & \propto
    \mbox{Re}\big[(\wilson[]{P} + \wilson[]{P'})\, \wilson[\ast]{T5}
                + (\wilson[]{S} + \wilson[]{S'})\, \wilson[\ast]{T}\big]
                \Big/ \Gamma_\ell \,,
\end{align}
i.e., with an interchange of tensor contributions $T \leftrightarrow T5$.  We
note also that $J_{6c}$ contributes to the lepton forward-backward asymmetry of
$B\to K^* \bar\ell\ell$ being $\propto (J_{6s} + J_{6c}/2)$. Since it has to
compete with $J_{6s}$ in this observable, a separate measurement of $J_{6s}$ and
$J_{6c}$ is necessary.

Only tensor contributions enter
\begin{align}
  (J_{1s} - 3\, J_{2s}) & \propto
  \left(\ldots |\wilson{T}|^2 + \ldots |\wilson{T5}|^2 \right) \,,
\end{align}
where the dots indicate different kinematic and form-factor dependencies.  But
tensor \emph{and} scalar contributions enter
\begin{equation}
\begin{aligned}
  (J_{1c} + J_{2c}) & \propto
      \ldots \left(|\wilson{T}|^2 + |\wilson{T5}|^2 \right)
\\
  & + \ldots \left(|\wilson{S} - \wilson{S'}|^2 + |\wilson{P} - \wilson{P'}|^2 \right) \,,
\end{aligned}
\end{equation}
which is similar to the dependence of $F_H^\ell$ in $B\to K
\bar\ell\ell$
\begin{equation}
  \label{eq:FH-dependence}
\begin{aligned}
  F_H^\ell  \propto & \Big[
      \ldots \left(|\wilson{T}|^2 + |\wilson{T5}|^2\right)
\\
  & + \ldots \left(|\wilson{S} + \wilson{S'}|^2 + |\wilson{P} + \wilson{P'}|^2 \right)
  \Big]
  \Big/ \Gamma_\ell \,.
\end{aligned}
\end{equation}
Concerning $F_H^\ell$, the involved kinematic factors---see
\cite{Bobeth:2007dw, Bobeth:2012vn}---are such that tensor and scalar
couplings contribute only constructively/cumulatively, apart from
cancellations among $\wilson{S(P)}$ and $\wilson{S'(P')}$.
Interference terms in the numerator of $F_H^\ell$ of the form
$(\wilson{T} \times \wilson{7,7',9,9'})$ and $(\wilson{P,P'} \times
\wilson{10,10'})$ are suppressed by $m_\ell/\sqrt{q^2}$. They become
numerically relevant in case $\wilson{T} \ll \wilson{7,7',9,9'}$ or
$\wilson{P,P'} \ll\wilson{10,10'}$ where the smallness of
$\wilson{T,P,P'}$ is of the same level as the suppression factor
$m_\ell/\sqrt{q^2}$ accompanying the large vectorial SM Wilson
coefficients $\wilson[SM]{9,10} \sim \pm 4$. This implies, however, no
large enhancement of $F_H^\ell$ over the SM prediction.

On the one hand, the observables $F_H^\ell$ \refeq{eq:FH-dependence} and $A_{\rm
  FB}^\ell$ \refeq{eq:AFB-dependence} are measured in the angular distribution
\refeq{eq:ang-distr} of $B\to K \bar\ell\ell$ normalized to the decay width
$\Gamma_\ell$ such that uncertainties due to form factors can cancel in part
\cite{Bobeth:2007dw, Bobeth:2012vn}. On the other hand, $J_{6c}$, $(J_{1s} - 3\,
J_{2s})$, and $(J_{1c} + J_{2c})$ appear in the unnormalized angular
distribution of $B\to K^* \bar\ell\ell$. ``Optimized'' versions $S_1$, $M_1$,
and $M_2$ for the low-$q^2$ region for which form factors cancel in the limit of
$m_b \to \infty$ have been identified in~\cite{Matias:2012xw}.  For the
high-$q^2$ region, potential normalizations are discussed in
\refapp{app:BKstarellell}, which could serve to form optimized observables for
special scenarios of either vanishing chirality-flipped vector or tensor or
scalar couplings. In the most general case, however, there are no optimized
observables at high $q^2$. Although form factors do not cancel in this case, it
might still be preferable to use normalizations, for example when the overall
normalization of $B\to K^*$ form factors constitutes a major theoretical
uncertainty.

\begin{table*}
\begin{center}
\renewcommand{\arraystretch}{1.4}
\begin{tabular}{lccc}
\hline
Channel & Constraints & Kinematics & Source
\\
\hline \hline
$B_s\to \bar\mu\mu$
& $\overline{\cal B} \equiv \int\dd\tau \BR(\tau)$ &  --  & \cite{Aaij:2013aka,Chatrchyan:2013bka,CMS:2014xfa}
\\
\hline
\multirow{5}{*}{${B^+\to K^+ \bar\mu\mu}$}
    & \multirow{2}{*}{$\BR_\mu$}
    & $q^2\in [1,\, 6],\, [14.18,\, 16],\, [> 16]$ GeV$^2$  & \cite{CDF:2012:BKstarll} \\
    &
    &  $\,\,\,\,\, q^2 \in [1.1,\, 6.0],\, [15.0,\, 22.0]$ GeV$^2$  & \cite{Aaij:2014pli} \\
    & $A_{\rm FB}^\mu$
    &  $\,\,\,\,\, q^2 \in [1.1,\, 6.0],\, [15.0,\, 22.0]$ GeV$^2$  & \cite{Aaij:2014tfa, CDF:2012:BKstarll} \\
    & $F_{H}^\mu$
    &  $\,\,\,\,\, q^2 \in [1.1,\, 6.0],\, [15.0,\, 22.0]$ GeV$^2$  & \cite{Aaij:2014tfa} \\
    & $A_{\rm CP}^\mu$
    &  $\,\,\,\,\, q^2 \in [1.1,\, 6.0],\, [15.0,\, 22.0]$ GeV$^2$  & \cite{Aaij:2014bsa}
\\
\hline
    \multirow{3}{*}{${B^0\to K^{\ast 0} \bar\mu\mu}$}
    & $\BR_\mu$
    & $q^2\in [1,\, 6],\, [14.18,\, 16],\, [> 16]$ GeV$^2$  & \cite{CDF:2012:BKstarll, Aaij:2013iag,Chatrchyan:2013cda} \\
    & $A_{\rm FB}^\mu$
    & $q^2\in [1,\, 6],\, [14.18,\, 16],\, [> 16]$ GeV$^2$  & \cite{CDF:2012:BKstarll, Aaij:2013iag,Chatrchyan:2013cda} \\
    & $A_{\rm CP}^\mu$
    &  $\,\,\,\,\, q^2 \in [1.1,\, 6.0],\, [15.0,\, 22.0]$ GeV$^2$  & \cite{Aaij:2014bsa}
\\
\hline
${B\to K}$ form factors
    & $f_{0,+,T}$   &  $q^2 = 17,\, 20,\, 23$ GeV$^2$  & \cite{Bouchard:2013eph} \\
\hline
\multirow{2}{*}{${B\to K^*}$ {form factors}}
    & $V,\,A_{0,1,2},\, T_{1,2,3}$ &  $q^2 = 0.1,\, 4.1,\, 8.1,\, 12.1$ GeV$^2$  & \cite{Straub:2015ica} \\
    & $V,\,A_{0,1,2},\, T_{1,2,3}$
    & $q^2 \in [11.9,\, 17.8]$ GeV$^2$  & \cite{Horgan:2013hoa, Horgan:2015vla}  \\
\hline
\end{tabular}
\renewcommand{\arraystretch}{1.0}
\caption{\label{tab:observables} List of all observables of the various $b\to s
  \bar\mu\mu$ decays entering the fits with the respective kinematics and
  experiments that provide the measurements. LCSR and lattice results of $B\to
  K^{(*)}$ form factors are used to constrain a $q^2$-dependent form-factor
  parametrization.  For more details see \refsec{sec:observables} and
  \refapp{app:theory:inputs}. \checked{} }
\end{center}
\end{table*}

To illustrate the sensitivity of $F_H^\ell$ to tensor couplings, we compare it
in \reffig{fig:FHvsCT} to the branching ratios of $B\to K^{(*)} \bar\ell\ell$
for {$\ell = \mu$}, integrated over one low-$q^2$ and one high-$q^2$ bin. The
details of the numerical input and the uncertainty propagation can be found in
\refapp{app:theory:inputs} and \refapp{app:mc}.  In light of the hint of new
physics in $\wilson{9}$ from recent global analyses of $b\to s (\gamma,\,
\bar\ell\ell)$ data \cite{Descotes-Genon:2013wba, Altmannshofer:2013foa,
  Beaujean:2013soa, Hurth:2014vma, Altmannshofer:2014rta}, we show predictions
for $\wilson[NP]{9} = -1.1$ in addition to $\wilson[NP]{9} = 0$.

From \reffig{fig:FHvsCT}, the highest sensitivity to tensor couplings of any $B
\to K$ observable is attained by {$F_H^\mu$} at high $q^2$ due to a partial
cancellation of form factors \cite{Bobeth:2012vn}. If the experimental
uncertainty could be reduced further, {$F_H^\mu$} would give a very strong
constraint on a simultaneous negative shift in $\wilson[NP]{9}$ and
{$\wilson{T}$}. The prediction of {${\cal B}(B\to K \bar\mu\mu)$} is
essentially insensitive to $\wilson{T}$ but sensitive to {$\wilson[NP]{9}$}. A
stronger impact on global fits, however, would {require a} reduced theory
uncertainty.

The observable {${\cal B}(B\to K^* \bar\mu\mu)$} shows moderate dependence on
$\wilson{T}$ at least at low $q^2$ and has some impact on the constraints on
tensor couplings as will be discussed in \refsec{sec:results}. At the moment,
theory and experimental uncertainty are of similar size.

$(J_{1c}+J_{2c})$ is sensitive to $\wilson{T}$ in both $q^2$ regimes. At low~$q^2$,
it is mildly affected by $\wilson[NP]{9}$, whereas at high $q^2$ it is
unaffected. Regarding $(J_{1s}-3J_{2s})$, the situation is reversed: here the
strong dependence on $\wilson{T}$ appears at low $q^2$.  Overall, {$F_H^\mu$},
$(J_{1c}+J_{2c})$ at {high} $q^2$, and $(J_{1s}-3J_{2s})$ are sensitive
to $\wilson{T}$ and theoretically very clean around $\wilson{T}=0$.

From the available measurements, {$F_H^\mu$} at high $q^2$ currently provides the
most stringent constraints on the size of tensor couplings. Moreover, the
dependence on vector couplings is such that $\wilson[NP]{9} \lesssim 0$ leads to
stronger constraints on~$\wilson{T}$ than $\wilson[NP]{9} \approx
0$.

Important additional constraints on scalar couplings come from the
branching ratio of $B_s \to \bar\mu\mu$ as given in
\refeq{eq:Br-Bsmumu}. It provides the most stringent constraints on
the moduli $|\wilson{S} - \wilson{S'}|$ and $|\wilson{P} -
\wilson{P'}|$ and further depends only on $(\wilson{10} -
\wilson{10'})$. Thus it is complementary to $F_H^{\ell}$ in $B\to K
\bar\ell\ell$; see Eq. \refeq{eq:FH-dependence}.

Eventually we also explore the effect of interference with {NP contributions}
in the vector couplings $\wilson{9,\,9',\,10,\,10'}$ on the bounds
on tensor and scalar couplings.  For this purpose we include also the branching
ratio, the lepton forward-backward asymmetry, and the rate CP asymmetry of
$B\to K^* \bar\mu\mu$ as they provide additional constraints on the real and
imaginary parts of $\wilson{9,\,9',\,10,\,10'}$. The experimental {input} of all
observables entering our fits is listed in \reftab{tab:observables} together with input for
the $B\to K^{(*)}$ form factors. More details on the latter can be found in
\refapp{app:theory:inputs}.

%
%
%
\section{
  \checked{Fits and constraints}
  \label{sec:results}
}

There are no discrepancies between the latest measurements for $\ell = \mu$
(throughout this section) of $F_H^\mu$ and $A_{\rm FB}^\mu$ in $B^+ \to K^+
\bar\mu\mu$ and their tiny SM predictions; cf. \reffig{fig:FHvsCT}. Thus our
main objective is to derive constraints on tensor and scalar couplings through
the enhanced sensitivity of both observables to {these} couplings compared to
vector couplings. For this purpose, we will consider several model-independent
scenarios, progressing from rather restricted to more general ones in order to
asses the effect of cancellations due to interference of various contributions.

For each coupling {that} we vary in a fit, we remain as general as possible, treat it
as a complex number and use the Cartesian parametrization assuming uniform
priors for ease of comparison with previous studies. Specifically, we set
\begin{align}
  \label{eq:wilson-prior}
  \mbox{Re}(\wilson{S,S',P,P',T,T5}) &\in [-1, 1],\nn \mbox{Re}(\wilson{9,9',10,10'}) &\in [-7, 7],
\end{align}
and the same for the imaginary parts. The priors of the nuisance parameters are given in \refapp{app:theory:inputs}.

We start with the scenario of only tensor couplings and see that they are well
constrained by $F_H^\mu$ alone. In a second scenario we consider only scalar
couplings in order to investigate the complementarity of $F_H^\mu$ and $B_s \to
\bar\mu\mu$. Here we find that---for the first time---all complex-valued scalar
couplings can be bounded simultaneously by the combination of both
measurements. Finally we consider as a special scenario the SM augmented by
dimension six operators as an effective theory of new physics below some high
scale $\Lambda_{\rm NP}$ assumed much larger than the typical scale of
electroweak symmetry breaking. In addition, the model contains one scalar
doublet under $SU(2)_L$ as in the SM. For each scenario, we also investigate
interference effects with new physics in vector couplings {$\wilson{9,9',10,10'}$}. Finally, we conclude
this section with posterior predictions---conditional on all experimental
constraints---of the probable ranges of the not-yet-measured angular observables
{$J_{6c}$, $(J_{1c}+J_{2c})$ and $(J_{1s} - 3J_{2s}$)}
in $B\to K^*\bar\mu\mu$ and $A_{\Delta\Gamma}$ in $B_s\to\bar\mu\mu$.

%
%
\subsection{Tensor couplings \label{sec:tensor}}

\begin{table*}
  \renewcommand{\arraystretch}{1.3}
  \begin{center}
  \begin{tabular}{cccc}
  \hline
  data set
  & only $F_H^\mu$
  & $F_H^\mu$ + other
  & $F_H^\mu$ + other
  \\
  set of couplings
  & $\wilson{T,\,T5}$
  & $\wilson{T,\,T5}$
  & $\wilson{T,\,T5,\, 9,\,10}$
  \\
  credibility level
  & 68\%, 95\%
  & 68\%, 95\%
  & 68\%, 95\%
  \\
  \hline
  $\mbox{Re}\,\wilson{T}$
  & $\quad$ $[-0.32,\, 0.16]$, $[-0.52,\, 0.35]$ $\quad$
  & $\quad$ $[-0.29,\, 0.09]$, $[-0.43,\, 0.27]$ $\quad$
  & $\quad$ $[-0.23,\, 0.14]$, $[-0.39,\, 0.30]$ $\quad$
  \\
  $\mbox{Im}\,\wilson{T}$
  & $[-0.25,\, 0.24]$, $[-0.44,\, 0.44]$
  & $[-0.19,\, 0.21]$, $[-0.37,\, 0.35]$
  & $[-0.17,\, 0.22]$, $[-0.33,\, 0.37]$
  \\[0.1cm]
  $\mbox{Re}\,\wilson{T5}$
  & $[-0.25,\, 0.24]$, $[-0.44,\, 0.44]$
  & $[-0.19,\, 0.17]$, $[-0.33,\, 0.33]$
  & $[-0.18,\, 0.16]$, $[-0.32,\, 0.32]$
  \\
  $\mbox{Im}\,\wilson{T5}$
  & $[-0.24,\, 0.25]$, $[-0.44,\, 0.45]$
  & $[-0.21,\, 0.19]$, $[-0.37,\, 0.35]$
  & $[-0.20,\, 0.18]$, $[-0.36,\, 0.35]$
  \\
  \hline
  $|\wilson{T}|$, $|\wilson{T5}|$
  & $[0.13,\, 0.43]$, $[0.03,\, 0.57]$
  & $[0.09,\, 0.33]$, $[0.02,\, 0.44]$
  & $[0.10,\, 0.33]$, $[0.02,\, 0.43]$
  \\
  \hline
  \end{tabular}

  \end{center}

  \renewcommand{\arraystretch}{1.0}
  \caption{
     \label{tab:cTT5:1-dimCLs}
     The constraints on complex-valued $\wilson{T,\,T5}$ when using
     measurements of only $F_H^\mu$, $F_H^\mu$ and other data in
     \reftab{tab:observables} (except $\overline{\cal B}(B_s \to \bar\mu\mu)$), and finally allowing
     complex-valued new physics contributions to $\wilson{9,\,10}$. \checked{} 
   }
\end{table*}

In a scenario with only {complex-valued tensor couplings} $\wilson{T,\, T5}$, the
experimental measurement of $F_H^\mu$ constrains the combination $|\wilson{T}|^2
+ |\wilson{T5}|^2$, up to some small interference of $\wilson{T}$ with vector
couplings $\wilson{7,\,7',\,9,\,9'}$; cf. \refeq{eq:FH-dependence}.  The
according 68\% (95\%) 1D-marginalized probability intervals are listed in the
second column of \reftab{tab:cTT5:1-dimCLs}. From the third column, it is seen
that the constraints become tighter when utilizing all observables in
\reftab{tab:observables}, mainly due to the sensitivity of the branching ratio
of $B\to K^{*} \bar\mu\mu$ to tensor couplings (see also
\reffig{fig:FHvsCT}). The latter stronger bounds are driven by the new lattice
results of $B\to K^*$ form factors that predict values above the measured
{ones}~\cite{Horgan:2013pva}. Since tensor couplings contribute constructively
to ${\cal B}(B\to K^{*}\bar\mu\mu)$, large values are better constrained. In
this scenario with vanishing scalar couplings, current measurements of $A_{\rm
  FB}^\mu(B^+\to K^+\bar\mu\mu)$ barely provide any constraint;
cf.~\refeq{eq:AFB-dependence}.

We also perform a fit with nonzero $\wilson[NP]{9,\,10}$ in order to assess the
robustness of the bounds with respect to interference. Note that $\wilson{7,7'}$
appears in linear combinations with $\wilson{9,9'}$ such that its interference
with tensor and scalar couplings is captured implicitly by allowing new physics
in $\wilson{9,9'}$. Thus we fix $\wilson{7,7'}$ to the SM value without loss of
generality. In this case, $F_H^\mu$ by itself still provides bounds on
$|\wilson{T,\,T5}|$ that are weakened by a factor of two since $F_H^\mu$ does
not pose constraints on $\wilson{9,\,10}$ (in the chosen prior range). Once
additional experimental measurements of \reftab{tab:observables} are taken into
account, the potential destructive effects of new physics in $\wilson{9,\,10}$
become reduced and almost the same constraints on $\wilson{T,\,T5}$ are
recovered, as shown in the last column in \reftab{tab:cTT5:1-dimCLs}. If in
addition we allow $\wilson{S,S',P,P'} \ne 0$ (not shown in
\reftab{tab:cTT5:1-dimCLs}), the credible regions further shrink by about 10\%,
which we attribute to {the} cumulative effect of $\wilson{S,S',P,P'} \ne 0$ in
$F_H^\ell$; cf. \refeq{eq:FH-dependence}. In summary, the $F_H^\mu$
measurement~\cite{Aaij:2014tfa} of LHCb with 3 fb$^{-1}$ shrinks the previous
bounds \cite{Bobeth:2012vn} on $\wilson{T,\,T5}$ by roughly 50\%.

A keen observer may notice that in \reftab{tab:cTT5:1-dimCLs} the SM point
$\wilson{T,T5} \equiv 0$ is contained in every 68\% region in Cartesian
coordinates but never in even the 95\% region in polar coordinates\footnote{But
  the bin with lower edge $\wilson{T,T5} = 0$ is always in the 99\%
  region.}. This is a consequence of the general \emph{concentration of
  measure}. Another way to look at it is to transform the uniform prior density
from Cartesian to polar coordinates. For the example of a single Wilson
coefficient, say $\wilson{T}$, the transformed density is proportional to the
determinant of the Jacobian which is $|\wilson{T}|$. Since the Cartesian prior
boundaries are much larger than the regions of high likelihood, one could think
the value of the boundary is irrelevant, but in fact it determines the peak of
the prior in polar coordinates. In other words, the uniform prior on
$\mbox{Re}(\wilson{T})$ and $\mbox{Im}(\wilson{T})$ favors larger values of
$|\wilson{T}|$ even though we consider it consensus in the community that
smaller rather than larger values are reasonable because $\wilson{T} = 0$ in the
SM. We suggest therefore that the default treatment be revised in the future to
include available prior knowledge.

%
%
\subsection{Scalar couplings \label{sec:scalar}}

\begin{figure*}
  \begin{center}
    \includegraphics[width=.325\textwidth]{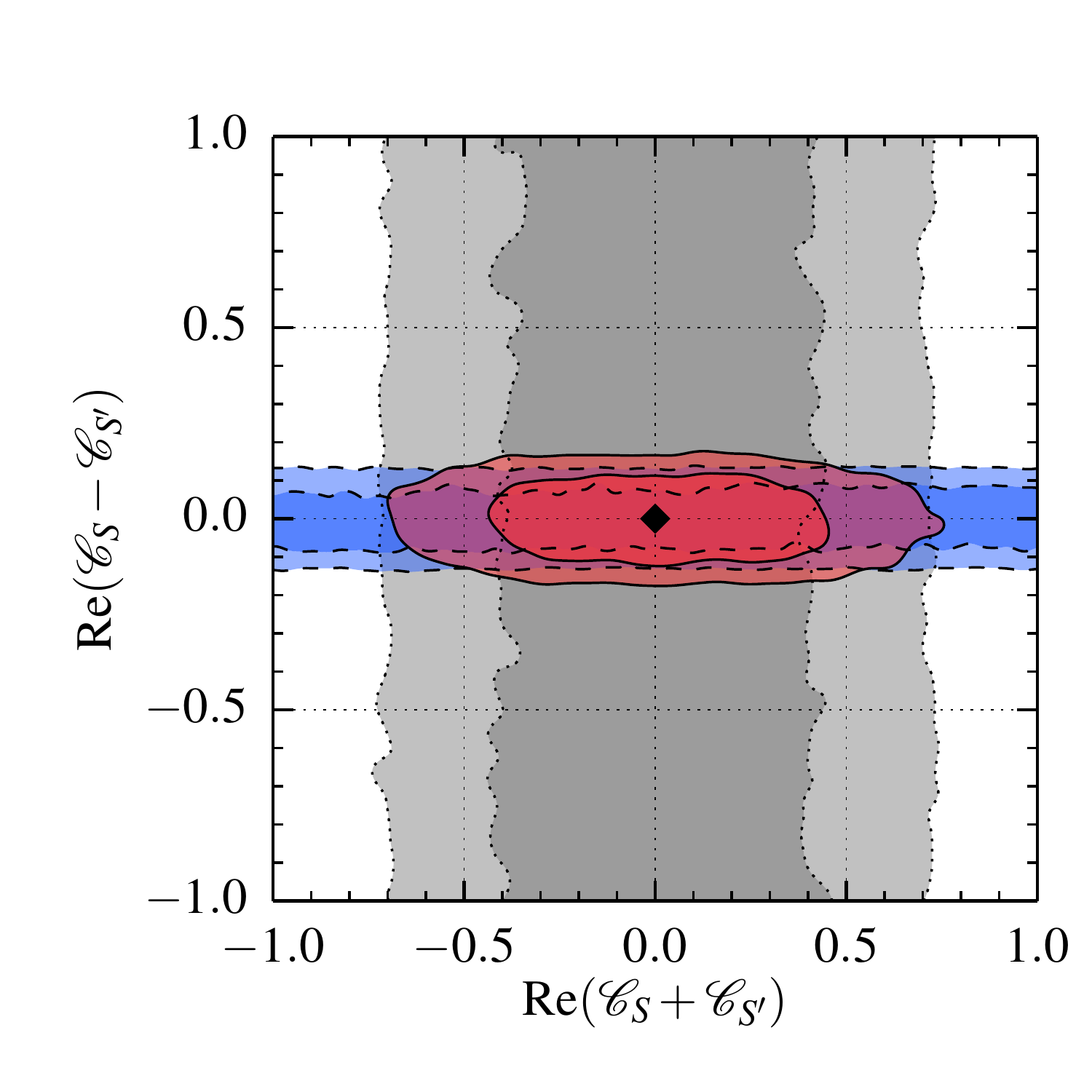}
    \includegraphics[width=.325\textwidth]{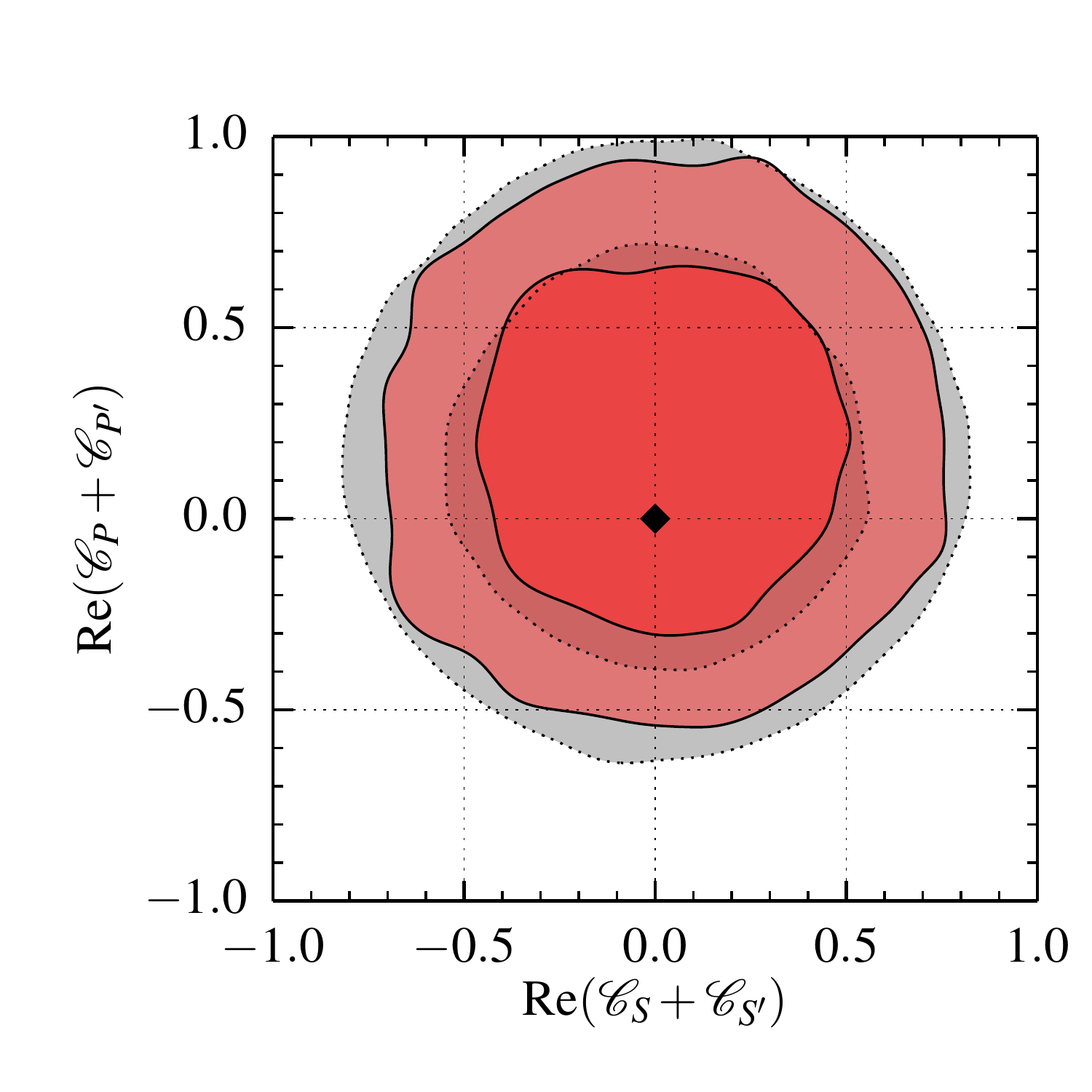}
    \includegraphics[width=.325\textwidth]{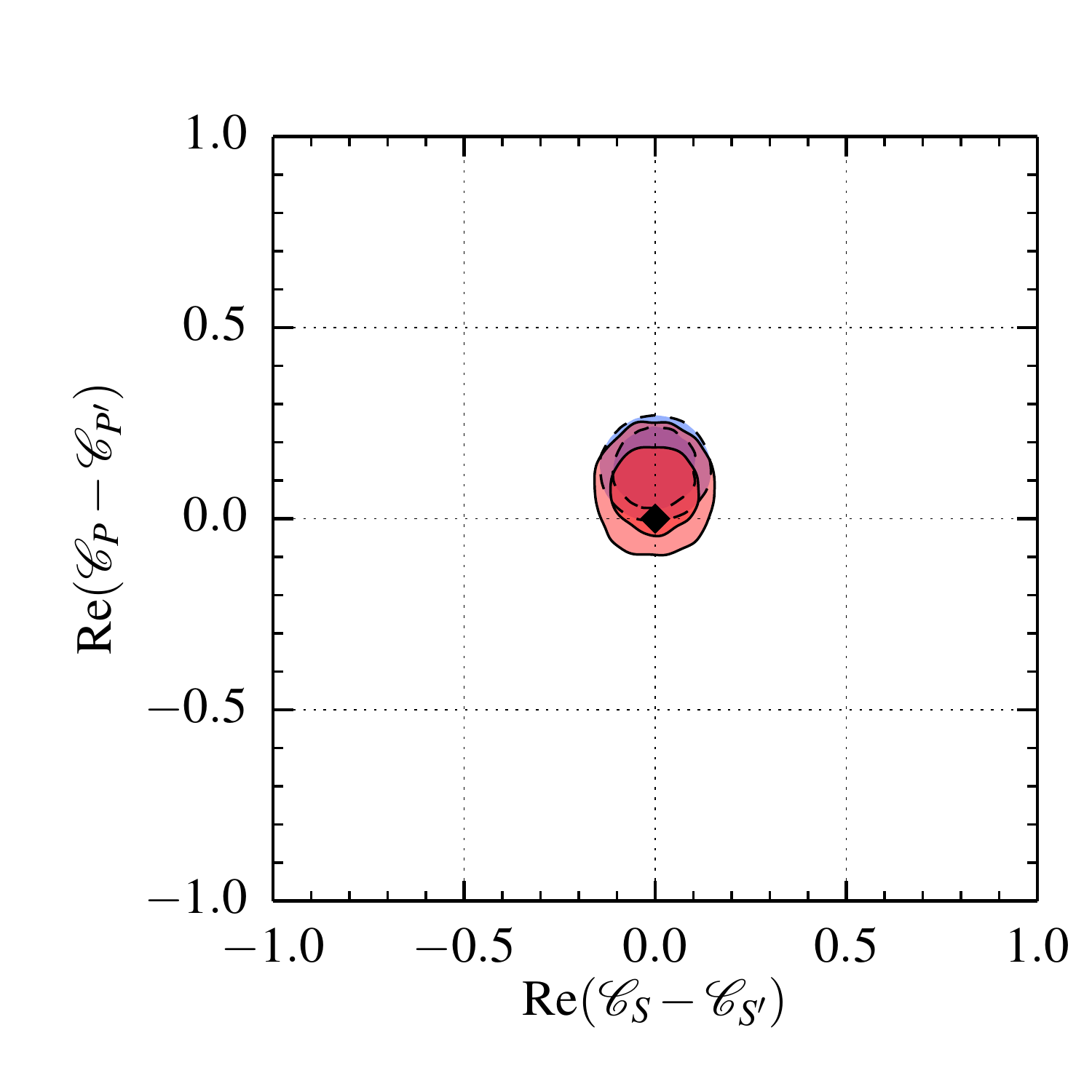}
  \end{center}
  \caption{ The constraints on complex-valued
    $\wilson{S,\,S',\,P,\,P'}$ from only $F_H^\mu$~(gray dotted), only
    $\overline{\cal B}(B_s \to \bar\mu\mu)$ (blue dashed), and the
    combination with all other data in \reftab{tab:observables} as
    well as nonzero $\wilson[]{9,\,9',\,10,\, 10'}$ (red solid) at
    68\% (darker) and 95\% (lighter) probability. The constraints on
    Re$(\wilson{P} \pm \wilson{P'})$ are identical to Re$(\wilson{S}
    \pm \wilson{S'})$, apart from a small translation of the contours
    by $(+0.2,\, +0.15)$. The SM prediction is indicated by the black
    diamond. }
  \label{fig:scalar}
\end{figure*}

\begin{table*}
  \renewcommand{\arraystretch}{1.3}
  \begin{center}
  \begin{tabular}{cccc}
  \hline
  data set
  & only $F_H^\mu$
  & only $\overline{\cal B}(B_s \to \bar\mu\mu)$
  & all
  \\
  set of couplings
  & $\wilson{S,\,S',\,P,\,P'}$
  & $\wilson{S,\,S',\,P,\,P'}$
  & $\wilson{S,\,S',\,P,\,P',\,9,\,9',\,10,\,10'}$
  \\
  credibility level
  & 68\%, 95\%
  & 68\%, 95\%
  & 68\%, 95\%
  \\
  \hline
  $\mbox{Re}(\wilson{S} - \wilson{S'})$
  & $\quad$ $-$ $\quad$
  & $\quad$ $[-0.10,\, 0.08]$, $[-0.14,\, 0.13]$ $\quad$
  & $\quad$ $[-0.08,\, 0.07]$, $[-0.13,\, 0.13]$ $\quad$
  \\
  $\mbox{Im}(\wilson{S} - \wilson{S'})$
  & $\quad$ $-$ $\quad$
  & $\quad$ $[-0.07,\, 0.07]$, $[-0.11,\, 0.12]$ $\quad$
  & $\quad$ $[-0.07,\, 0.07]$, $[-0.12,\, 0.11]$ $\quad$
  \\[0.1cm]
  $\mbox{Re}(\wilson{S} + \wilson{S'})$
  & $[-0.36,\, 0.39]$, $[-0.69,\, 0.68]$
  & $-$
  & $[-0.32,\, 0.32]$, $[-0.59,\, 0.62]$
  \\
  $\mbox{Im}(\wilson{S} + \wilson{S'})$
  & $[-0.37,\, 0.35]$, $[-0.68,\, 0.66]$
  & $-$
  & $[-0.25,\, 0.41]$, $[-0.57,\, 0.64]$
  \\[0.1cm]
  $\mbox{Re}(\wilson{P} - \wilson{P'})$
  & $-$
  & $[ 0.05,\, 0.20]$, $[ 0.01,\, 0.26]$
  & $[ 0.00,\, 0.16]$, $[-0.07,\, 0.22]$ \hspace{-3mm}
  \\
  $\mbox{Im}(\wilson{P} - \wilson{P'})$
  & $-$
  & $[-0.07,\, 0.08]$, $[-0.12,\, 0.12]$
  & $[-0.07,\, 0.09]$, $[-0.14,\, 0.16]$
  \\[0.1cm]
  $\mbox{Re}(\wilson{P} + \wilson{P'})$
  & $[-0.24,\, 0.51]$, $[-0.51,\, 0.82]$
  & $-$
  & $[-0.12,\, 0.52]$, $[-0.42,\, 0.78]$
  \\
  $\mbox{Im}(\wilson{P} + \wilson{P'})$
  & $[-0.36,\, 0.37]$, $[-0.67,\, 0.67]$
  & $-$
  & $[-0.37,\, 0.29]$, $[-0.68,\, 0.57]$
  \\
  \hline
  \end{tabular}
  \end{center}

  \renewcommand{\arraystretch}{1.0}
  \caption{
     \label{tab:cSP:1-dimCLs}
     The 1D-marginalized constraints on complex-valued $\wilson{S,\,S',\,P,\,P'}$
     at 68\% (95\%) probability from measurements of only $F_H^\mu$,
     only $\overline{\cal B}(B_s \to \bar\mu\mu)$, and all the data in
     \reftab{tab:observables} and additional new physics contributions to
     $\wilson{9,\, 9',\, 10,\,10'}$. \checked{}
   }
\end{table*}

Scalar couplings $\wilson{S,\, S',\, P,\, P'}$ enter $F_H^\ell$
without kinematic suppression---see \refeq{eq:FH-dependence}---as the
sum $(\wilson{i} + \wilson{i'})$ whereas in the time-integrated
branching ratio $\overline{\cal B}(B_s\to \bar\ell\ell)$ they appear
as the difference $(\wilson{i} - \wilson{i'}), i = S,\, P$.  Since the
existing measurement of $F_H^\mu$ constrains the sum, the combination
of $F_H^\mu$ and $\overline{\cal B}(B_s \to \bar\mu\mu)$ allows
us---for the first time---to bound the real and imaginary parts of all
four couplings.  The corresponding 2D-marginalized regions in the
$\mbox{Re}(\wilson{i} \pm \wilson{i'})$ ($i = S, P$) planes are shown
in \reffig{fig:scalar}. The corresponding plots for
$\mbox{Im}(\wilson{i} \pm \wilson{i'})$ are very similar to those
shown and thus omitted. These bounds do not change when including all
other data in \reftab{tab:observables}, since the $A_{\rm FB}^\mu(B\to
K\bar\mu\mu)$ requires interference of scalar with tensor couplings
and other observables are not very sensitive to scalar
couplings. Quantitatively, the constraint from $\overline{\cal B} (B_s
\to \bar\mu\mu)$ on $(\wilson{i} - \wilson{i'})$ is about a factor
four to five stronger than the one of $F_H^\mu$ on $(\wilson{i} +
\wilson{i'})$.

Interference terms of $\wilson{P,\,P'}$ with vector couplings might weaken these
bounds. For $\overline{\cal B}(B_s \to \bar\mu\mu)$, the relevant term is
$(\wilson{10} - \wilson{10'})$ (see \refeq{eq:Br-Bsmumu:time0}) and for
$F_H^\ell$ it is $(\wilson{10} + \wilson{10'})$ \cite{Bobeth:2007dw}; both are
suppressed by the factors $m_\mu/M_{B_s}$ and $m_\mu/\sqrt{q^2}$,
respectively. Nevertheless, these terms become important for small
$\wilson{P,\,P'}$ due to the large SM value of $\wilson[SM]{10} \simeq -4.2$. We
compile bounds on complex-valued scalar couplings in \reftab{tab:cSP:1-dimCLs}
for only $F_H^\mu$, only $\overline{\cal B}(B_s \to \bar\mu\mu)$, and their
combination with all the other observables in \reftab{tab:observables}. Neither
$F_H^\ell$ nor $\overline{\cal B}(B_s \to \bar\ell\ell)$ alone can bound all
four complex-valued scalar couplings, however their combination is capable to do
so and moreover, the bounds are stable against destructive interference with
vector couplings. In the case of $\wilson[NP]{9,9',10,10'} \ne 0$, we find the
following bounds with 68\% (95\%) probability:
\begin{equation}
  \label{eq:SP-bounds}
  \begin{aligned}
    \left| \wilson{S} \right| & \in [0.06,\, 0.25]\; ([0.01,\, 0.36]),\\
    \left| \wilson{S'} \right| & \in [0.09,\, 0.46]\; ([0.02,\, 0.73]),\\
    \left| \wilson{P} \right| & \in [0.05,\, 0.21]\; ([0.01,\, 0.32]),\\
    \left| \wilson{P'} \right| & \in [0.11,\, 0.42]\; ([0.02,\, 0.63]).
  \end{aligned}
\end{equation}
Allowing in addition $\wilson{T,T5} \ne 0$, the intervals of
\refeq{eq:SP-bounds} are quite similar but in general (10--20)\% narrower and
shifted by that amount towards zero. Again, this can be explained by the
cumulative effect of $\wilson{S,S',P,P'}$ and $\wilson{T,T5}$ in $F_H^\ell$
shown in \refeq{eq:FH-dependence}.

In the special case of real-valued couplings, $\overline{\cal B}(B_s \to
\bar\ell\ell)$ would lead to rings~\cite{Alonso:2014csa} instead of circles in \reffig{fig:scalar}.
Our results improve and extend previous bounds in the literature to the most
general case of complex-valued couplings. For example they are a factor two to
five more stringent than~\cite{Becirevic:2012fy} and comparable to
\cite{Altmannshofer:2012az} once restricting to the simpler scenarios considered
there.
%
%
\subsection{SM-EFT-constrained scalar couplings \label{sec:SM-EFT}}

In the following we consider a scenario in which it is assumed that there is
a sizable hierarchy between the electroweak scale and the new-physics
scale, $\Lambda_{\rm NP}$, and that the SM gauge symmetries $SU(2)_L \times
U(1)_Y$ are only broken at the electroweak scale. This results in the
augmentation of the SM by dimension-six operators that respect the SM gauge
group and are composed of SM fields only. Such a scenario becomes more
and more viable for two reasons. The first is the discovery of a
scalar resonance at the LHC in agreement with all requirements of the
Higgs particle in the SM. The second is the steadily rising lower
bound on the mass of new particles reported by ATLAS and CMS in
various more or less specific models.

A nonredundant set of dimension-six operators of this effective theory (SM-EFT)
that requires a linear realization of the electroweak symmetry was given in
\cite{Grzadkowski:2010es}.  The matching of the SM-EFT to the effective theory
of $\Delta B = 1$ decays \refeq{eq:Heff} at the scale $\mu \sim m_W$ of the
order of the $W$-boson mass was performed for vector couplings
$\wilson{7,\,9,\,10}$ in~\cite{D'Ambrosio:2002ex}. The matching of tensor
\refeq{eq:tensor:ops} and scalar \refeq{eq:scalar:ops} operators
\cite{Alonso:2014csa} shows that SM gauge groups in conjunction with the linear
representation impose the relations
\begin{align}
  \label{eq:mEFT-constraints}
  \wilson{P}  & = -\wilson{S}\,, &
  \wilson{P'} & = \wilson{S'}\,, &
  \wilson{T}  & = \wilson{T5} = 0
\end{align}
on scalar couplings and require tensor couplings to be
suppressed to the level of dimension-eight operators. In consequence only two
scalar couplings $\wilson{S,\, S'}$ arise that scale as $(\sim v/\Lambda_{\rm
  NP})^2 \ll 1$ where $v \propto m_W$ denotes the scale of electroweak symmetry breaking.

It must be noted that the relations \refeq{eq:mEFT-constraints} are a
consequence of embedding the Higgs in a weak doublet along with the Goldstone
bosons.  For example, choosing a nonlinear representation of the scalar sector
allows additional dimension-six operators in the according effective theory,
such that the couplings $\wilson{S,S',P,P'}$ are all independent and tensor
operators have nonvanishing couplings already at dimension six
\cite{Cata:2015lta}.

Omitting for the sake of simplicity terms of order $m_\ell^2/M_{B_s}^2$ and
$m_s/m_b$, the couplings $\wilson{S,\, S'}$ can be bound from~\cite{Alonso:2014csa}
\begin{equation}
\begin{aligned}
  \overline{\cal B}(B_s\to \bar\ell\ell) & \propto
      \left|\wilson{S} + \wilson{S'}\right|^2
    + \left|\wilson{S} - \wilson{S'}\right|^2
\\[0.2cm]
  - & \frac{4 m_\ell\, m_b}{M_{B_s}^2} \mbox{Re}
  \left[ (\wilson{S} + \wilson{S'})(\wilson{10} - \wilson{10'})^* \right] \,.
\end{aligned}
\end{equation}
In similar spirit, dropping terms of order $m_\ell^2/q^2$ and
$m_s/m_b$ gives
\begin{equation}
\begin{aligned}
  F_H^\ell & \propto
      \left|\wilson{S} + \wilson{S'}\right|^2
    + \left|\wilson{S} - \wilson{S'}\right|^2
\\[0.2cm]
  & - \frac{4 m_\ell\, m_b}{q^2} \mbox{Re}
  \left[ (\wilson{S} - \wilson{S'})(\wilson{10} + \wilson{10'})^* \right] \,.
\end{aligned}
\end{equation}

\begin{figure*}
  \begin{center}
      \includegraphics[width=.35\textwidth]{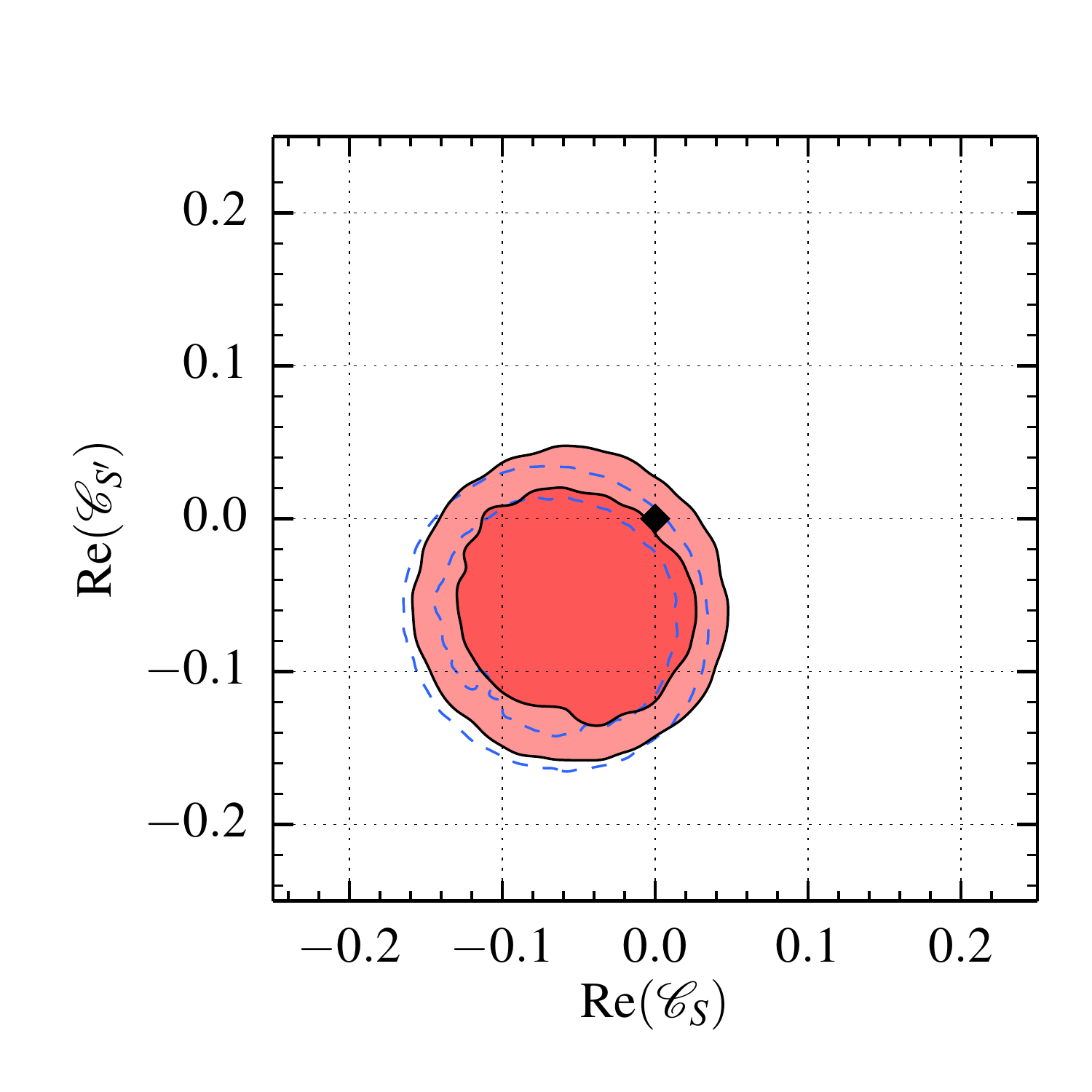}
      \includegraphics[width=.35\textwidth]{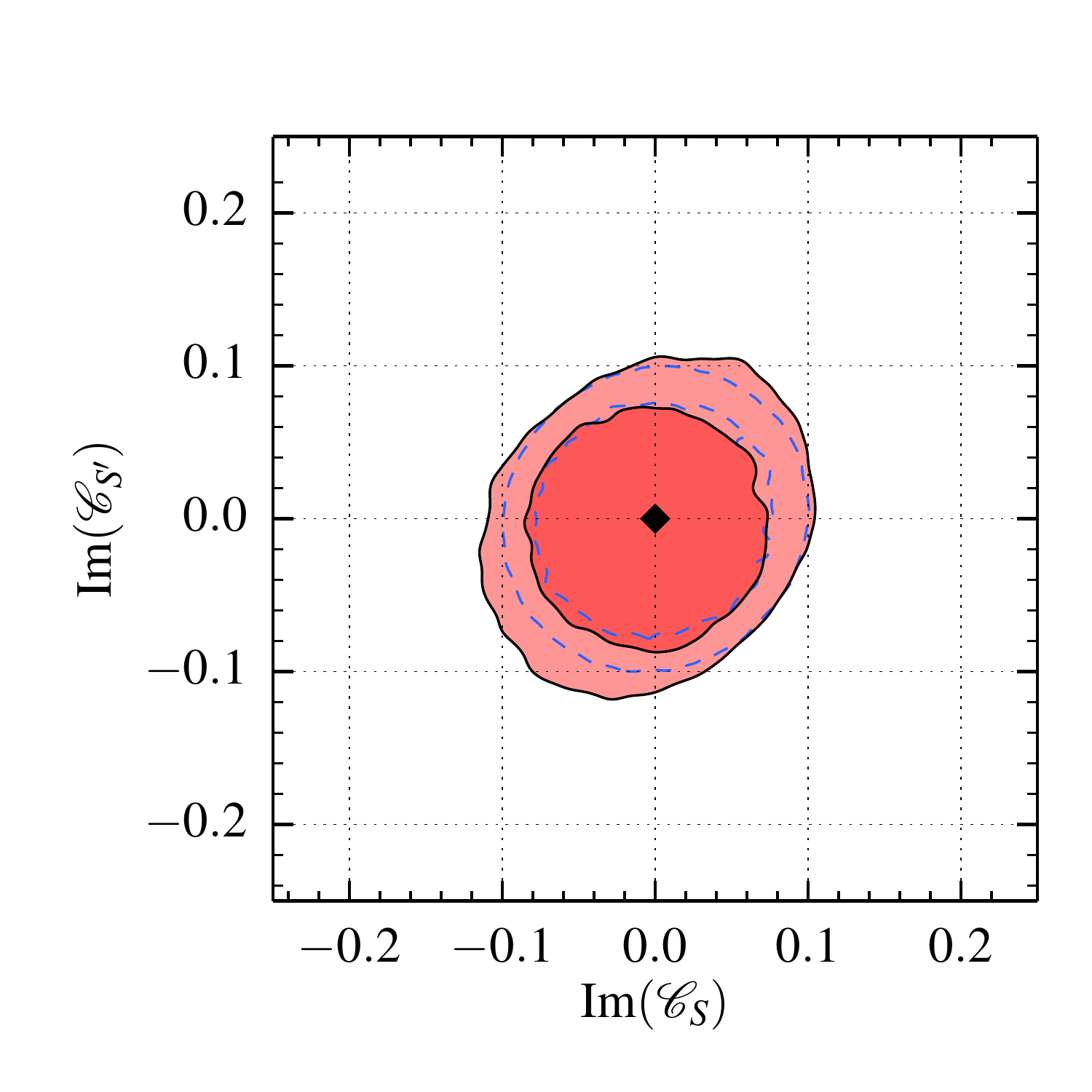}
    \end{center}
    \caption{ 68\% and 95\% contours of the 2D-marginalized distributions of
      scalar couplings $\wilson{S,\,S'}$ in the scenario SM-EFT with all
      constraints in \reftab{tab:observables} (red) when marginalizing over
      nonzero $\mbox{Re, Im} (\wilson[NP]{10,\,10'})$.  For comparison, we
      superimpose the corresponding contours from using only $\overline{\cal
        B}(B_s\to \bar\mu\mu)$ and all {$B \to K \bar\mu\mu$} constraints
      (dashed blue contours) with fixed $\wilson[NP]{10,\,10'} = 0$.  The SM is
      indicated by the black diamond.  \checked{}}
  \label{fig:scalar:SM-EFT}
\end{figure*}

In the SM-EFT no relations between $\wilson{10}$ and$\wilson{10'}$ arise,
so they are in general additional independent parameters. Here
we find that destructive interference with contributions involving
$\wilson{10,\, 10'}$ does not significantly alter the bounds on $\wilson{S,\,
  S'}$.  The results of two fits are shown in \reffig{fig:scalar:SM-EFT}. In the
first fit, we set $\wilson[NP]{10, 10'} = 0$ and include all constraints on
{$B \to K \bar\mu\mu$ and $B_s \to \bar\mu\mu$}. In the second fit, we allow
$\wilson[NP]{10, 10'} \ne 0$ and further include all {$B \to K^* \bar\mu\mu$}
constraints from \reftab{tab:observables}. For both fits, all six 2D marginals
of real and imaginary parts of $\wilson{S}$ vs. $\wilson{S'}$ have nearly
circular contours of equal size that contain the SM point at the 68\% level
except for $\mbox{Re}(\wilson{S})$ vs. $\mbox{Re}(\wilson{S'})$ where it is
within the 95\% credible region. The regions hardly vary between the two fits.

Since we consider here complex-valued couplings the allowed regions are circles
rather than rings as for the case of real-valued
couplings~\cite{Alonso:2014csa}. Compared to those rings, the circles are
smaller because the probability moves from the ring towards the center of the
circle.

%
%
\subsection{Tensor, scalar, and vector couplings \label{sec:tensor:scalar}}

The most general fit of complex-valued tensor and scalar couplings
$\wilson{S,S',P,P',T,T5}$ in combination with vector couplings
$\wilson{9,\,9',\,10,\, 10'}$---the combination of \refsec{sec:tensor} and
\refsec{sec:scalar}---yields bounds very similar to those in tables
\ref{tab:cTT5:1-dimCLs} and~\ref{tab:cSP:1-dimCLs}. The changes are only small
and in fact the bounds tend to be even more stringent because tensor and scalar
couplings can contribute to $F_H^{\mu}$ only constructively---see
\refeq{eq:FH-dependence}.  As before, branching-ratio measurements of $B\to
K^{*}\bar\mu\mu$ help to improve the constraints on $\wilson{T,\, T5}$. This
demonstrates that even in the case of complex-valued couplings there is enough
information in the data to bound all 20 real and imaginary parts.

%
%
\subsection{Angular observables in $B\to K^* \bar\ell\ell$}

\begin{table*}
  \renewcommand{\arraystretch}{1.5}
  \begin{center}
  \begin{tabular}{cccccc}
  \hline
    observable
  & $q^2$-bin [GeV$^2$]
  & SM
  & $\quad T(5),\,9,\,10\quad$
  & $\quad S^{(')},\,P^{(')},\,9^{(')},\,10^{(')} \quad$
  & $\quad S^{(')},P^{(')},\,T(5),\,9^{(')},10^{(')}$
  \\
  \hline
  \multirow{2}{*}{$\tau_{B^0} \times J_{6c}$}
  & $[1.1,\, 6]$
  & $\simeq 0$
  & $(0.6_{-1.9}^{+1.8}) \cdot 10^{-9\;}$
  & $(-0.1_{-1.9}^{+2.3}) \cdot 10^{-10}$
  & $(0.2_{-1.1}^{+1.1}) \cdot 10^{-9}$
  \\
  & $[15,\, 19]$
  & $\simeq 0$
  & $(2.1_{-6.2}^{+5.1}) \cdot 10^{-10}$
  & $(0.7_{-6.6}^{+5.4}) \cdot 10^{-11}$
  & $(-0.2_{-2.9}^{+3.5}) \cdot 10^{-11}$
  \\
  \hline
  \multirow{2}{*}{$\tau_{B^0} \times (J_{1c} + J_{2c})$}
  & $[1.1,\, 6]$
  & $(3.30_{-0.56}^{+0.65}) \cdot 10^{-9\;}$
  & $(4.8_{-1.9}^{+1.7}) \cdot 10^{-9\;}$
  & $(2.6_{-0.4}^{+0.5}) \cdot 10^{-9\;}$
  & $(2.8_{-1.1}^{+1.5}) \cdot 10^{-9}$
  \\
  & $[15,\, 19]$
  & $(1.72_{-0.16}^{+0.16}) \cdot 10^{-10}$
  & $(4.7_{-2.6}^{+3.3}) \cdot 10^{-9\;}$
  & $(1.9_{-0.7}^{+0.3}) \cdot 10^{-10}$
  & $(3.1_{-2.3}^{+2.5}) \cdot 10^{-9}$
  \\
  \hline
  \multirow{2}{*}{$\tau_{B^0} \times (J_{1s} - 3 J_{2s})$}
  & $[1.1,\, 6]$
  & $(2.44_{-0.48}^{+0.46}) \cdot 10^{-10}$
  & $(1.9_{-1.1}^{+1.4}) \cdot 10^{-8\;}$
  & $(4.0_{-1.2}^{+1.6}) \cdot 10^{-10}$
  & $(1.5_{-1.1}^{+1.0}) \cdot 10^{-8}$
  \\
  & $[15,\, 19]$
  & $(1.12_{-0.10}^{+0.10}) \cdot 10^{-10}$
  & $(7.1_{-3.8}^{+5.6}) \cdot 10^{-9\;}$
  & $(1.4_{-0.4}^{+0.3}) \cdot 10^{-10}$
  & $(5.3_{-3.8}^{+3.8}) \cdot 10^{-9}$
  \\
  \hline
  $A_{\Delta \Gamma}(B_s \to \bar\mu\mu)$
  & --
  & 1
  & 1
  & $[-1,\, 1]$
  & $[-1,\, 1]$
  \\
  \hline
  \end{tabular}
  \end{center}
  \renewcommand{\arraystretch}{1.0}
  \caption{
     \label{tab:postdictions}
     The posterior predictive 68\% probability intervals of not-yet-measured
     angular observables for several new-physics scenarios given all the
     considered experimental constraints. The corresponding values for the
     SM (prior predictive) are given, too, where ``$\simeq 0$'' indicates
     zero in the considered approximation, see text for details.
   }
\end{table*}

Now we discuss what the {fits tell} us about likely values of observables that
have not been measured yet but have sensitivity to tensor and scalar couplings.
In the $B\to K^*(\to K\pi)\bar\ell\ell$ decay, we again consider $(J_{1s} - 3\,
J_{2s})$ and $(J_{1c} + J_{2c})$ as in \refsec{sec:observables} and additionally
$J_{6c}$. We compute the posterior predictive distribution (see ~\ref{app:mc})
for each observable integrated over the low-$q^2$ bin $[1.1,\, 6]$~GeV$^2$ and
high-$q^2$ bin $[15,\, 19]$~GeV$^2$ matching LHCb's range. The distributions
resemble Gaussians, thus we summarize them by their modes and smallest 68\%
intervals in \reftab{tab:postdictions} comparing the SM (prior predictive,
$\wilson[NP]{i}=0$) to three NP scenarios. In each, we allow for interference
with the vector couplings and additionally vary only $\wilson{T,T5}$
(\refsec{sec:tensor}), only $\wilson{S,S',P,P'}$ (\refsec{sec:scalar}),
and finally both tensor and scalar couplings (\refsec{sec:tensor:scalar}).

We rescale $J_i$ and combinations by the $B^0$-meson life time
$\tau_{B^0} = 1.519$~ps~\cite{Agashe:2014kda} to judge the
experimental sensitivity in the near future by comparing to current
measurements of the branching ratio
\begin{align}
  {\BR} &
  = \frac{\tau_{B^0}}{3} \big[
    2\, (3 J_{1s} - J_{2s}) + (3 J_{1c} -  J_{2c})
  \big] \,.
\end{align}
In the SM, the typical magnitude of the branching ratio of $B\to
K^*\bar\ell\ell$ is approximately equal to $2 \cdot 10^{-7}$ for both
the {$q^2 \in [1.1,\, 6]$} and $[15,\, 19]$ GeV$^2$
bins. For comparison, the predicted ranges for $\tau_{B^0}(J_{1s} -
3\, J_{2s})$ and $\tau_{B^0}(J_{1c} + J_{2c})$ in the SM are
suppressed by $2-3$ orders of magnitude down to ${\cal O}(10^{-10})$;
cf. \reftab{tab:postdictions}.

The angular observable $J_{6c}$ is strictly zero in the absence of tensor and
scalar couplings. Nonzero contributions can be generated in the SM by QED
corrections or potentially from higher-dimensional ($d \geq 8$) $|\Delta B| =
|\Delta S| = 1$ operators, leading to parametric suppression by $\alpha_e/(4
\pi)$ or {$m_b m_\ell/m_W^2$}. These factors should be compared to the potential
suppression present for tensor and scalar contributions in particular NP models
in order to gauge their relevance. Our model-independent fits are still in a
regime where such considerations are insignificant since current experimental
measurements, in combination with theory uncertainties, do not yet impose
{sufficiently}
stringent constraints on tensor and scalar couplings.

Beyond the SM, $J_{6c}$ can become of order ${\cal O}(10^{-9})$ in scenarios
involving tensor couplings only and about ${\cal O}(10^{-10})$ in the presence
of scalar couplings only. Both effects are due to interference with vector
couplings.  Schematically, $J_{6c}$ is a function of $\wilson{T} \times
\wilson{P} + \wilson{S} \times \wilson{T5}$, {$m_{\ell}/\sqrt{q^2} \times
\mbox{scalar} \times \mbox{vector}$, and $m_{\ell}/\sqrt{q^2} \times
\mbox{tensor} \times \mbox{vector}$}. The largest interval is obtained for
the scenario without scalar couplings because then the uncertainty on the
(tensor) couplings is largest. But even then, it {seems that the} experimental
sensitivity will not be high enough to have an impact in global fits.

Concerning $(J_{1c} + J_{2c})$ and $(J_{1s} - 3\, J_{2s})$, substantial
deviations from the SM prediction are again only possible in the presence of
nonzero tensor couplings. In this case, an enhancement by two orders of
magnitude is possible up to ${\cal O}(10^{-8})$ at high $q^2$ and also at low
$q^2$ in the case of $(J_{1s} - 3\, J_{2s})$. We want to stress again that we
make these statements conditional on all included experimental constraints, the
scenario, and our prior. In view of the current experimental precision of $20\%$
on the branching ratio at LHCb~\cite{Aaij:2013iag} with only 1 fb$^{-1}$,
corresponding to the ${\cal O}(10^{-8})$, one can indeed hope for some
sensitivity to such large effects in $(J_{1c} + J_{2c})$ and $(J_{1s} - 3\,
J_{2s})$ for the not-yet-published 3 fb$^{-1}$ data set. At least, we can hope
for some measurement if the method of moments~\cite{beaujean:2015mom} is
applied.

For the CP asymmetry $A_{\Delta\Gamma}(B_s\to \bar\mu\mu)$ induced by
the nonvanishing width of the $B_s$ meson
(cf. \refapp{app:BKstarellell}), we find a rather uniform distribution
in scenarios with nonzero scalar couplings. So any value in the range
$[-1,\, 1]$ is plausible whereas {the SM and the scenario
  with only tensor couplings predict a value of precisely one
  \cite{DeBruyn:2012wk}}; cf. the last row in
\reftab{tab:postdictions}. Hence any deviation from one would
unambiguously hint at the presence of scalar operators.

%
%
%
\section{\checked{Conclusions}}

We have derived the most stringent constraints to date on tensor and scalar
couplings that mediate $b\to s\bar\mu\mu$ transitions{. They are based on}
the latest
measurements of angular observables $F_H^\mu$ and the lepton forward-backward
asymmetry $A_{\rm FB}^\mu$ in $B^+\to K^+ \bar\mu\mu$ from LHCb
\cite{Aaij:2014tfa}, supplemented by measurements of the branching ratios of
$B_s\to \bar\mu\mu$ and $B\to K^{(*)} \bar\mu\mu$.

Both $F_H^\mu$ and $A_{\rm FB}^\mu$ belong to a class of observables
in which vector and dipole couplings---present in the standard model
(SM)---are suppressed (mostly kinematically by $m_\ell / \sqrt{q^2}$)
with respect to tensor and scalar couplings. We provide predictions
for the equivalent but not-yet-measured angular observables $J_{6c}$,
$(J_{1c} + J_{2c})$ and $(J_{1s} - 3 J_{2s})$ in $B\to K^* (\to K\pi)
\bar\mu\mu$.

In a Bayesian analysis of the complex-valued couplings of the effective theory,
we find that the measurement of $F_H^\mu$, especially at high-$q^2$,
\begin{enumerate}
\item imposes by itself {constraints} on tensor couplings
  $|\wilson{T,\,T5}|$ such that the upper bound of the smallest 68\%
  (95\%) credibility interval is $0.43\, (0.57)$, superseding previous
  bounds.  In combination with current data from $B\to K^* \bar\mu\mu$
  and lattice {predictions of} $B\to K^*$ form factors, the
  bounds are lowered to 0.33 (0.43), even in the presence of
  nonstandard contributions in vector couplings.
  \item for the first time allows to {simultaneously} bound all four scalar couplings
  {$\wilson{S,S',P,P'}$} due to its complementarity to
  $\overline{\cal B}(B_s \to \bar\mu\mu)$. Even when taking into account
  destructive interference with vector couplings, $|\wilson{i} + \wilson{i'}| <
  0.3\, (0.6)$ and $|\wilson{i} - \wilson{i'}| < 0.1\, (0.2)$ for $i = S,P$ with
  at least 68\% (95\%) probability. Currently, the bounds from $F_H^\mu$ are
  weaker than those from $\overline{\cal B}(B_s \to \bar\mu\mu)$ by about a
  factor of four. Future measurements of $F_H^\mu$ at LHCb and Belle II will
  further tighten the bounds.

  Moreover, measurements of $F_H^e$ ($\ell = e$) will provide constraints on
  scalar couplings in the electron channel in the absence of a direct
  determination of the branching ratio of $B_s \to \bar{e}e$.
\end{enumerate}
Our updated bounds on complex-valued tensor and scalar couplings are summarized
in \reftab{tab:cTT5:1-dimCLs} and \reftab{tab:cSP:1-dimCLs}, accounting also for
interference effects with vector couplings. These bounds hold even in the most
general scenario of complex-valued tensor, scalar, and vector couplings, showing
that the data are good enough to bound the real and imaginary parts of all
Wilson coefficients simultaneously.

As a special case, we consider the scenario arising from the SM
augmented by dimension-6 operators generalizing existing studies to
the case of complex-valued couplings. In this scenario, tensor
couplings are absent and additional relations between scalar couplings
are enforced by the linear realization of the $SU(2)_L \otimes U(1)_Y$
electroweak symmetry group.

Our study of the {yet unmeasured} angular observables $J_{6c}$,
$(J_{1c} + J_{2c})$, and $(J_{1s} - 3 J_{2s})$ in $B\to K^* (\to K\pi)
\bar\mu\mu$ (see \reftab{tab:postdictions}) shows that despite the
current bounds on tensor couplings, enhancements of up to two orders
of magnitude over the SM predictions are allowed for $(J_{1c} +
J_{2c})$ and $(J_{1s} - 3 J_{2s})$, placing them in reach of the LHCb
analysis of the full run~I data set. Our bounds on scalar couplings
from $B_s\to \bar\mu\mu$ and $F_H^\mu$, however, are already quite
restrictive permitting only small deviations from SM predictions in
$(J_{1c} + J_{2c})$ and $(J_{1s} - 3 J_{2s})$. Notably, the CP
asymmetry $A_{\Delta\Gamma}(B_s\to \bar\mu\mu)$ given nonzero scalar
couplings can take on any value in the range $[-1,\, 1]$.

%
%
%
\begin{acknowledgements}
  We are grateful to Danny van Dyk for helpful discussions and his support on
  \eos~\cite{EOS}. Concerning form factor results from lattice QCD, we thank
  Chris Bouchard and Matthew Wingate for communications on their results of
  $B\to K$ \cite{Bouchard:2013eph} and $B\to K^*$ \cite{Horgan:2015vla} form
  factors. We thank also David Straub for his support with LCSR results of $B\to
  K^*$ form factors \cite{Straub:2015ica}. We thank Martin Jung, David Straub,
  and Danny van Dyk for comments on the manuscript. C.B. was supported by the ERC
  Advanced Grant project ``FLAVOUR'' (267104). We acknowledge the support by the
  DFG Cluster of Excellence "Origin and Structure of the Universe". The
  computations have been carried out on the computing facilities of the
  Computational Center for Particle and Astrophysics (C2PAP).
\end{acknowledgements}

%
%

\appendix

%
%
%
\section{
  \checked{Angular observables in $B\to K^* \bar\ell\ell$}
  \label{app:BKstarellell}
}

Here we focus on those angular observables $J_i$ in $B\to K^* \bar\ell\ell$ in
which tensor and scalar contributions are kinematically enhanced by a factor
$\sqrt{q^2}/m_\ell$ compared to the vector contributions of the SM or the
respective interference terms. The results and the notation follow
\cite{Bobeth:2012vn}.

For convenience, we split the time-like transversity amplitude $A_t$ into two
parts as
\begin{align}
  A_t &
  = \tilde{A}_t - \frac{1}{2} \frac{\sqrt{q^2}}{m_\ell} A_P \, .
\end{align}
Both terms depend on the scalar $B\to K^*$ form factor $A_0(q^2)$, a
normalization factor $N$, and the K{\"a}ll{\'e}n-function $\lambda$
(see~\cite{Bobeth:2012vn}), but have a different dependence on the
Wilson coefficients $i = 10, 10', P, P'$:
\begin{equation}
\begin{aligned}
  \tilde{A}_t & = 2 N \frac{\sqrt{\lambda}}{\sqrt{q^2}}
    (\wilson[]{10} - \wilson[]{10'}) A_0\,,
\\
  A_P & = - 2 N\sqrt{\lambda} \frac{(\wilson[]{P} - \wilson[]{P'})}{(m_b+m_s)} A_0\,.
\end{aligned}
\end{equation}
The lepton-flavor index $\ell$ of Wilson coefficients is omitted for brevity
throughout.

In full generality, the angular observables in $B\to K^* \bar\ell\ell$ depend on
seven transversity amplitudes with vector and dipole contributions,
$A_{0,\,\perp,\,\parallel}^{L,R}$ and $\tilde{A}_t$, one scalar and one
pseudoscalar amplitude, $A_{S,\,P} \propto (\wilson[]{S,P} - \wilson[]{S',P'})$,
and six tensor amplitudes, $A_{\parallel\perp,\,t\perp,\,0\parallel} \propto
\wilson[]{T}$ and $A_{t0,\,t\parallel,\,0\perp} \propto \wilson[]{T5}$.  The
interesting combinations are
\begin{align}
  \frac{4}{3} J_{6c}  =
  4\, \beta_\ell\, \mbox{Re} \bigg[ &2\, (A_{t0}^{} A_S^* - A_{\parallel\perp}^{} A_P^*) \nonumber\\
  &+ \frac{m_\ell}{\sqrt{q^2}} \big[(A_0^L + A_0^R) A_S^*
           + 4\, A_{\parallel\perp}^{} \tilde{A}_t^* \big] \bigg],
\end{align}
\begin{align}
  \label{eq:j1c-j2c}
  \frac{4}{3}\(J_{1c} + J_{2c}\) = \;
  &\Bigg| \frac{2\, m_\ell}{\sqrt{q^2}} (A_0^L + A_0^R) + 4\, A_{t0} \Bigg|^2    + 16\, \beta_\ell^2\, |A_{\parallel\perp}|^2 \nonumber\\
  + &\Bigg|\frac{2\, m_\ell}{\sqrt{q^2}}\, \tilde{A}_t - A_P\Bigg|^2 +
  \beta_\ell^2\, |A_S|^2,
 \end{align}
 \begin{align}
 \frac{4}{3}\(J_{1s} - 3\, J_{2s}\) = \;
    &\Bigg| \frac{\sqrt{2}\, m_\ell}{\sqrt{q^2}} (A_\perp^L + A_\perp^R) + 4\, A_{t\perp} \Bigg|^2 \nonumber\\
  + &\Bigg| \frac{\sqrt{2}\, m_\ell}{\sqrt{q^2}} (A_\parallel^L + A_\parallel^R) + 4\, A_{t\parallel} \Bigg|^2.
\end{align}
The function $\beta_\ell^2(q^2) \equiv 1 - 4 m_\ell^2/q^2$ tends to 1 for $m_\ell \ll
\sqrt{q^2}$.  This condition is well fulfilled for $\ell = e$ and $q^2 \gtrsim
1\, \mbox{GeV}^2$, provided that tensor and scalar Wilson coefficients do not
receive additional suppression factors. For $\ell = \mu$, the value of $q^2$
should not be too low, whereas in the case $\ell = \tau$, these observables are
not anymore dominated by tensor and scalar contributions alone, and the full
lepton-mass dependence has to be taken into account.  Finally, we note that the
second part of $(J_{1c} + J_{2c})$ in \refeq{eq:j1c-j2c},
\begin{align}
  &\Bigg|\frac{2\, m_\ell}{\sqrt{q^2}}\, \tilde{A}_t - A_P\Bigg|^2
   + \beta_\ell^2\, |A_S|^2   =  4 N^2 \lambda A_0^2 \Bigg\{
    \beta_\ell^2 \Bigg|\frac{\wilson[]{S} - \wilson[']{S}}{m_b + m_s}\Bigg|^2\nonumber\\
  &\hskip1cm + \Bigg|\frac{\wilson[]{P} - \wilson[']{P}}{m_b + m_s}
     + \frac{2 m_\ell}{q^2} \big(\wilson[]{10} - \wilson[']{10}\big) \Bigg|^2
    \Bigg\}\,,
\end{align}
 resembles very much the branching ratio of the rare
decay $B_s \to \bar\ell\ell$ in the limit $q^2 \to M_{B_s}^2$
\begin{align}
\label{eq:Br-Bsmumu:time0}
  {\cal B}(B_s \to \bar\ell\ell) &=
    \frac{G_F^2 \alpha_e^2 |V_{tb}^{} V_{ts}^*|^2}{64\, \pi^3}
    M_{B_s}^5 f_{B_s}^2 \tau_{B_s} \beta_\ell(q^2 = M_{B_s}^2)
\nonumber\\
    & \times \Bigg\{
    \beta_\ell^2(q^2 = M_{B_s}^2) \Bigg|\frac{\wilson[]{S} - \wilson[']{S}}{m_b + m_s}\Bigg|^2 \nonumber\\
    &+ \Bigg|\frac{\wilson[]{P} - \wilson[']{P}}{m_b + m_s}
     + \frac{2 m_\ell}{M_{B_s}^2} \big(\wilson[]{10} - \wilson[']{10}\big) \Bigg|^2
    \Bigg\}\,.
  \end{align}
  Hence there is some similarity between $(J_{1c} + J_{2c})$ in $B \to K^*
  \bar\ell\ell$ and ${\cal B}(B_s \to \bar\ell\ell)$ in their dependence on the
  couplings but the former has additional dependence on tensor and vector
  couplings through the other transversity amplitudes.  In $B \to K^*
  \bar\ell\ell$, the helicity suppression factor $4\,m_\ell^2/q^2$ of vector
  couplings is weaker than the corresponding factor $4\,m_\ell^2/M_{B_s}^2$ in
  $B_s \to \bar\ell\ell$.

  Concerning $B_s\to \bar\ell\ell$, the expression \refeq{eq:Br-Bsmumu:time0}
  corresponds to the plain branching ratio at time $t=0$. Due to the
  nonvanishing decay width $\Delta\Gamma_s$, experiments measure the average
  time-integrated branching ratio---denoted by $\overline{\cal B}$---and the two
  are related as \cite{DeBruyn:2012wk}
\begin{align}
  \label{eq:Br-Bsmumu}
  \overline{\cal B} (B_s \to \bar\ell\ell) &
  = \frac{1 + y_s {\cal A}_{\Delta\Gamma}}{1 - y_s^2} {\cal B} (B_s \to \bar\ell\ell)\,.
\end{align}
Here $y_s \equiv \Delta\Gamma_s/(2\Gamma_s)$ with the numerical value given in
\cite{Beaujean:2013soa}. ${\cal A}_{\Delta\Gamma}$ is the CP asymmetry
due to nonvanishing width difference, which is ${\cal A}_{\Delta\Gamma} = 1$
in the SM, but in general can be ${\cal A}_{\Delta\Gamma} \in [-1,\, 1]$.
Since ${\cal A}_{\Delta\Gamma}$ can depart from it's SM value in scenarios of
new physics considered in this work, we take this effect into account
in our numerical analysis, although it is suppressed by small $y_s$. The latest
SM prediction $\overline{\cal B} (B_s \to \bar\mu\mu) = (3.65 \pm 0.23) \cdot
10^{-9}$~\cite{Bobeth:2013uxa} includes NLO electroweak \cite{Bobeth:2013tba}
and NNLO QCD corrections \cite{Hermann:2013kca}.

Finally we discuss the possibility of suitable normalizations of $J_{6c}$,
$(J_{1s} - 3\, J_{2s})$ and $(J_{1c} + J_{2c})$ at high $q^2$ that would provide
optimized observables. For this purpose we use form-factor relations at leading
order in $1/m_b$ and neglect terms suppressed by $m_\ell/\sqrt{q^2}$. With the
notation and expressions derived in \cite{Bobeth:2012vn},
\begin{equation}
\begin{aligned}
  J_{1s} - 3\, J_{2s} & = \frac{3}{2}\, \rho_1^T (f_\perp^2 + f_\parallel^2) \,,
\\[0.1cm]
  J_{1c} + J_{2c} & = 3\, \rho_1^T f_0^2
     + \frac{3 N^2 \lambda}{(m_b+m_s)^2} A_0^2
\\
   & \quad \times \left(|\wilson{S}-\wilson{S'}|^2 + |\wilson{P}-\wilson{P'}|^2\right)\,.
\end{aligned}
\end{equation}
Here $f_{\perp, \parallel, 0}$ and $A_0$ denote $B\to K^*$ form factors, whereas
the $\rho_1^\pm$ depends on vector couplings and $\rho_1^T \propto
(|\wilson{T}|^2 + |\wilson{T5}|^2)$.

Concerning $(J_{1s} - 3\, J_{2s})$, there are no appropriate normalizations,
unless the chirality-flipped $\wilson[NP]{7',\,9',\,10'} = 0$ because then
$\rho_1^+ = \rho_1^-$ holds. There are three potential normalizations
\begin{equation}
\begin{aligned}
  \frac{4}{3} & J_{1s} =
    (3 \rho_1^+ + \rho_1^T) f_\perp^2 +  (3 \rho_1^- + \rho_1^T) f_\parallel^2 \,,
\\
  \frac{8}{3} & J_{2s} =
    (\rho_1^+ - \rho_1^T) f_\perp^2 +
    (\rho_1^- - \rho_1^T) f_\parallel^2 \,,
\\[0.1cm]
  & J_{1s} + 2 J_{2s} =
    3 (\rho_1^+ f_\perp^2 + \rho_1^- f_\parallel^2) \,
\end{aligned}
\end{equation}
where the last one depends only on vector couplings; i.e., is free of tensor and
scalar ones. In $J_{1s}$ and $J_{2s}$, tensor couplings contribute either
cumulatively or destructively to vector couplings in $\rho_1^\pm$.

A similar situation arises for $(J_{1c} + J_{2c})$, where no appropriate
normalization exists, unless scalar couplings vanish. In this special case, both
$J_{1c}$ and $J_{2c} = -3/2 (\rho_1^- - \rho_1^T) f_0^2$ depend only on $f_0$
and can be used as normalization. In the general case, $J_{2c}$ still depends
only on tensor couplings and was used at low $q^2$ \cite{Matias:2012xw}. Finally
we note that
\begin{equation}
\begin{aligned}
  J_{1c} - J_{2c} & = 3\,\rho_1^- f_0^2 + \frac{3 N^2 \lambda}{(m_b+m_s)^2} A_0^2
\\
   & \quad \times \left(|\wilson{S}-\wilson{S'}|^2 + |\wilson{P}-\wilson{P'}|^2\right)\,.
\end{aligned}
\end{equation}
is free of tensor couplings at high $q^2$ and would provide access to
$\rho_1^-$ provided scalar couplings vanish.

%
%
%
\section{
  \checked{Theoretical Inputs}
  \label{app:theory:inputs}
}

Here we describe the theoretical treatment of observables and collect the
numerical input for the relevant parameters. The software package \eos{}
\cite{Bobeth:2011gi, Bobeth:2011nj, EOS} is used for the calculation of
observables in $B_s\to \bar\mu\mu$ and $B\to K^{(*)}\bar\ell\ell$ and associated
constraints. Both the likelihood and the prior are defined entirely within \eos.

Concerning numerical input, we refer the reader to~\cite{Beaujean:2013soa} for
the compilation of nuisance parameters relevant to this work. We adopt the same
values for fixed parameters and the same priors unless noted otherwise
below. Specifically, we use identical priors for for common nuisance parameters
of the CKM quark-mixing matrix, the charm and bottom quark masses in the
$\overline{\rm MS}$ scheme, and the parametrization of subleading corrections
in $1/m_b$ as given in \cite{Beaujean:2013soa}.

Contrary to \cite{Beaujean:2013soa}, we do not choose a log-gamma
distribution for the asymmetric uncertainties in priors anymore but
rather a continuous yet asymmetric Gaussian distribution. This avoids
a poor fit because the log-gamma distribution falls off too rapidly in
the ``short'' tail. In a unifying spirit, we now use the continuous rather
than discontinuous asymmetric Gaussian to approximate asymmetric
experimental intervals in our likelihood.

The updated prior of the $B_s$ decay constant $f_{B_s}$ enters the branching
ratio of $B_s\to \bar\mu\mu$. We adopt recent updates of the $N_f = (2 + 1)$
FLAG compilation \cite{Aoki:2013ldr}, see \reftab{tab:hadronic:nuisance}, which
averages the results of \cite{Bazavov:2011aa, McNeile:2011ng, Na:2012kp}.  More
recent calculations with $N_f = (2+1+1)$ \cite{Dowdall:2013tga} and $N_f = 2$
\cite{Carrasco:2013zta} flavors are consistent with these averages.

The tensor and scalar amplitudes in $B\to K^{(*)} \bar\ell\ell$ factorize
naively, i.e. they depend only on scalar and tensor $B\to K^{(*)}$ form
factors. These amplitudes are implemented in \eos{} for $B\to K \bar\ell\ell$
and $B\to K^* \bar\ell\ell$ as given in \cite{Bobeth:2007dw} and
\cite{Bobeth:2012vn}, respectively and we refrain from using form-factor
relations at {low and high $q^2$} for the tensor and scalar $B\to K$ and $B\to
K^*$ form factors.

{Therefore, we need} additional nuisance parameters for the
complete set of $B\to K$ form factors $f_{+,T,0}$. As a consequence of using the
$z$ parametrization \cite{Khodjamirian:2010vf} and the kinematic relation
$f_0(0) = f_+(0)$, five nuisance parameters (listed in
\reftab{tab:hadronic:nuisance}) are needed. As prior information, we average the
LCSR results~\cite{Khodjamirian:2010vf} and \cite{Ball:2004ye} (see
\reftab{tab:hadronic:nuisance} and cf. \cite{Beaujean:2013soa}) supplement them
with lattice determinations~\cite{Bouchard:2013eph}.
The lattice results are given in a slightly different parametrization
necessitating a conversion to the parametrization used here. For this purpose,
we generate the form factors from the parametrization
of~\cite{Bouchard:2013eph}, including full correlations, at three values of $q^2
= 17,\, 20,\, 23$ GeV$^2$.  Subsequently, these ``data'' are included in the
prior by means of a multivariate Gaussian whose mean and covariance is given in
\reftab{tab:btok-lattice}. The $q^2$ values and number of points are chosen such
that the correlation of neighboring points is small enough to keep the
covariance nonsingular.

\begin{table}
\begin{center}
\renewcommand{\arraystretch}{1.4}
\begin{tabular}{cccc}
\hline
  Quantity & Prior & Reference\\
\hline
  \multicolumn{3}{c}{$B_s$ decay constant}
\\
\hline
  $f_{B_s}$   &  ($227.7 \pm 4.5$) MeV &  \cite{Aoki:2013ldr}\\
\hline
  \multicolumn{3}{c}{$B\to K$ form factors}\\
\hline
  $f_+(0)=f_0(0)$    &  $0.34 \pm 0.05$       &  \cite{Khodjamirian:2010vf, Ball:2004ye}\\
  $f_T(0)$    &  $0.38 \pm 0.06$      &  \cite{Khodjamirian:2010vf, Ball:2004ye}\\
  $b_1^0$     &  $-4.3^{+0.8}_{-0.9}$    &  \cite{Khodjamirian:2010vf}\\
  $b_1^+$     &  $-2.1^{+0.9}_{-1.6}$    &  \cite{Khodjamirian:2010vf}\\
  $b_1^T$     &  $-2.2^{+1.0}_{-2.0}$    &  \cite{Khodjamirian:2010vf}\\
\hline
\multicolumn{3}{c}{$B\to K^*$ form factors}\\
\hline
$\alpha_{0}^{A_0}$     & $0.35^{+0.02}_{-0.03}$      & \cite{Straub:2015ica, Horgan:2015vla}\\
$\alpha_{1}^{A_0}$     & $-1.21^{+0.08}_{-0.08}$     & \cite{Straub:2015ica, Horgan:2015vla}\\
$\alpha_{2}^{A_0}$     & $0.77^{+0.45}_{-0.48}$      & \cite{Straub:2015ica, Horgan:2015vla}\\
$\alpha_{0}^{A_1}$     & $0.2625^{+0.015}_{-0.015}$  & \cite{Straub:2015ica, Horgan:2015vla}\\
$\alpha_{1}^{A_1}$     & $0.07^{+0.08}_{-0.08}$      & \cite{Straub:2015ica, Horgan:2015vla}\\
$\alpha_{2}^{A_1}$     & $0.045^{+0.12}_{-0.15}$     & \cite{Straub:2015ica, Horgan:2015vla}\\
$\alpha_{1}^{A_{12}}$  & $0.53^{+0.10}_{-0.14}$     & \cite{Straub:2015ica, Horgan:2015vla}\\
$\alpha_{2}^{A_{12}}$  & $0.32^{+0.33}_{-0.39}$      & \cite{Straub:2015ica, Horgan:2015vla}\\
$\alpha_{0}^{V}$       & $0.35^{+0.02}_{-0.03}$      & \cite{Straub:2015ica, Horgan:2015vla}\\
$\alpha_{1}^{V}$       & $-1.19^{+0.08}_{-0.08}$     & \cite{Straub:2015ica, Horgan:2015vla}\\
$\alpha_{2}^{V}$       & $1.55^{+0.33}_{-0.33}$      & \cite{Straub:2015ica, Horgan:2015vla}\\
$\alpha_{0}^{T_1}$     & $0.31^{+0.02}_{-0.02}$      & \cite{Straub:2015ica, Horgan:2015vla}\\
$\alpha_{1}^{T_1}$     & $-1.07^{+0.08}_{-0.08}$     & \cite{Straub:2015ica, Horgan:2015vla}\\
$\alpha_{2}^{T_1}$     & $1.40^{+0.30}_{-0.30}$      & \cite{Straub:2015ica, Horgan:2015vla}\\
$\alpha_{1}^{T_2}$     & $0.33^{+0.08}_{-0.10}$     & \cite{Straub:2015ica, Horgan:2015vla}\\
$\alpha_{2}^{T_2}$     & $0.26^{+0.21}_{-0.24}$      & \cite{Straub:2015ica, Horgan:2015vla}\\
$\alpha_{0}^{T_{23}}$  & $0.67^{+0.04}_{-0.05}$      & \cite{Straub:2015ica, Horgan:2015vla}\\
$\alpha_{1}^{T_{23}}$  & $0.93^{+0.28}_{-0.22}$     & \cite{Straub:2015ica, Horgan:2015vla}\\
$\alpha_{2}^{T_{23}}$  & $-0.15^{+0.86}_{-0.71}$      & \cite{Straub:2015ica, Horgan:2015vla}\\
\hline
\end{tabular}
\renewcommand{\arraystretch}{1.0}
\caption{\label{tab:hadronic:nuisance} Prior distributions of the nuisance
  parameters for hadronic quantities. \checked{} }
\end{center}
\end{table}

\begin{table*}
  \begin{center}
    \renewcommand{\arraystretch}{1.3}
    \begin{tabular}{cccc}
        \hline
        $q^2$ [\GeV$^2$] & $17$            & $20$            & $23$\\
        \hline
        $f_0(q^2)$   & $0.62 \pm 0.03$ & $0.72 \pm 0.03$ & $0.87 \pm 0.04$\\
        $f_+(q^2)$   & $1.13 \pm 0.05$ & $1.63 \pm 0.07$ & $2.68 \pm 0.13$\\
        $f_T(q^2)$   & $1.02 \pm 0.06$ & $1.47 \pm 0.08$ & $2.42 \pm 0.18$\\
        \hline
    \end{tabular}\\
    \renewcommand{\arraystretch}{1.0}
    \vspace{2\smallskipamount}
    \renewcommand{\arraystretch}{1.3}
    \begin{tabular}{cccccccccc}
        \hline
        $q^2$ [\GeV$^2$]
             & $17$  & $20$ & $23$ & $17$ & $20$ & $23$ & $17$  & $20$   & $23$ \\
        \hline
   $f_0(17)$ & 1.00  & 0.93 & 0.73 & 0.58 & 0.48 & 0.23 & 0.13  & 0.010 & 0.048 \\
   $f_0(20)$ & --    & 1.00 & 0.87 & 0.41 & 0.45 & 0.24 & 0.059 & 0.093 & 0.075 \\
   $f_0(23)$ & --    & --   & 1.00 & 0.28 & 0.34 & 0.32 & 0.010 & 0.046 & 0.058 \\
   $f_+(17)$ & --    & --   & --   & 1.00 & 0.78 & 0.30 & 0.36  & 0.21  & 0.051 \\
   $f_+(20)$ & --    & --   & --   & --   & 1.00 & 0.71 & 0.22  & 0.26  & 0.20  \\
   $f_+(23)$ & --    & --   & --   & --   & --   & 1.00 &-0.025 & 0.14  & 0.22  \\
   $f_T(17)$ & --    & --   & --   & --   & --   & --   & 1.00  & 0.79  & 0.46  \\
   $f_T(20)$ & --    & --   & --   & --   & --   & --   & --    & 1.00  & 0.85  \\
   $f_T(23)$ & --    & --   & --   & --   & --   & --   & --    & --    & 1.00  \\
        \hline
    \end{tabular}

  \end{center}

    \renewcommand{\arraystretch}{1.0}
    \caption{Mean values, standard deviations (top) and correlation
        coefficients (bottom) of lattice points \cite{Bouchard:2013eph} for the
        $B\to K$ form factors $f_{0,+,T}(q^2)$. \checked{}
        \label{tab:btok-lattice}
      }
\end{table*}

We change the parametrization of $B\to K^*$ form factors w.r.t. our previous
work \cite{Beaujean:2013soa} slightly and now use the simplified series
expansion (SSE) \cite{Bharucha:2010im}
\begin{align}
  \label{eq:SSE-FF-parametrization}
  F_i(q^2) & =
  \frac{1}{1 - q^2/M_{R,i}^2} \sum_k \alpha_k^i \left[z(q^2) - z(0)\right]^k \,,
\end{align}
with three parameters $\alpha_k^i$ ($k=0,1,2$) per form factor $i = V,\, A_0,A_1,
T_1,T_{23}$ and two ($k=1,2$) for $i= A_{12}, T_2$. The parameters $\alpha_0^i$
correspond to $F_i(q^2 = 0)$ and the kinematic constraints
\begin{align}
  A_{12}(0) & = {\cal F}_{\rm kin}\, A_0(0) \,, &
  T_1(0) & = T_2(0) \,,
\end{align}
are used to eliminate $\alpha_0^{A_{12}, T_2}$, where the kinematic factor
${\cal F}_{\rm kin} \equiv (M_B^2 - M_{K^*}^2)/(8 M_B M_{K^*})$ depends on the
$B$- and $K^*$-meson masses. The {pole} masses $M_{R,i}$ are set to
the values given in \cite{Straub:2015ica}.

This parametrization allows us to consistently combine available results of
form factors from different nonperturbative approaches, namely LCSR's at large
recoil and lattice at low recoil and to simultaneously implement the kinematic
constraints. We determine the parameters $\alpha_i^k$ in a combined fit to
form-factor predictions of $V, A_{0,1,2},T_{1,2,3}$ from the LCSR
\cite{Straub:2015ica} and of $V, A_0,A_1,A_{12},T_1,T_2,T_{23}$ from the lattice
\cite{Horgan:2013hoa, Horgan:2015vla} calculations, including their
correlations. We verified that the  constraint
\begin{align}
  \label{eq:A12-constraint-q2max}
  A_{12}(q_{\rm max}^2) & = {\cal F}_{\rm kin}\, A_1(q_{\rm max}^2)
\end{align}
at the kinematic endpoint $q_{\rm max}^2 \equiv (M_B - M_{K^*})^2$ is satisfied
to high accuracy by the above constraints and thus need not be imposed
explicitly. Uninformative flat priors are chosen for all $\alpha_k^i$ with
ranges
\begin{align}
  \alpha_0^i & \in [0,\, 1] \,, &
  \alpha_{1}^i & \in [-2,\, 2]  \,, &
  \alpha_{2}^i & \in [-3,\, 3] \,.
\end{align}
The results of the LCSR form-factor predictions have been provided to us
directly by the authors of \cite{Straub:2015ica} at $q^2 = (0.1,\, 4.1,\, 8.1,\,
12.1)\, \mbox{GeV}^2$ for $V, A_0,A_1,A_{12},T_1,T_2,T_{3}$ including the
$28\times 28$ covariance matrix. Similar results could have been obtained by
``drawing'' form factors from the correlated parameters given in
\cite{Straub:2015ica} and ancillary files just as for $B \to K$ lattice form
factors.

For the $B \to K^*$ lattice form factors this approach is not good
enough as we were not able to select more than two $q^2$ values
without obtaining a singular covariance. In order to fully the exploit
the available information, we then contacted the authors
of~\cite{Horgan:2015vla} and obtained the original values of the form
factors (including correlation) at various values in the interval $q^2
\in [11.9,\, 17.8]$ to which Horgan et al. fit the SSE. The covariance
has a block-diagonal structure with a 48$\times$48 block for $V, A_0,
A_1, A_{12}$ and a 36$\times$36 block for $T_1, T_2, T_{23}$. Having
the ``raw'' information on form-factor values is much more reliable
and future proof as there are no issues with artificial correlation
and we could one day decide to use yet another form-factor
parametrization and fit it easily to these data points.

We have compared the SSE fit \refeq{eq:SSE-FF-parametrization} with two versus
three parameters and found that in the former case lattice form factors
influence the fit such that form factors tend to be higher than LCSR predictions
at low $q^2$ leading to a poor fit. Hence we prefer the three-parameter setup as
it provides the flexibility needed to accommodate LCSR and lattice results. Means
and standard {deviations} of that three-parameter fit are given in
\reftab{tab:hadronic:nuisance}, we omit the correlations for the sake of brevity
but are happy to provide them.

\section{\checked{Monte Carlo sampling} \label{app:mc}}

The marginalization of the posterior is performed
with the package {\tt pypmc}~\cite{beaujean_2015_20045}, which incorporates the
algorithm presented in \cite{Beaujean:2013:PMC, Beaujean:2012uj} and in addition
an implementation of the variational Bayes algorithm. In every analysis we first
run multiple adaptive Markov chains (MCMC) in parallel through
\texttt{pypmc}. If necessary, chains are seeded at the SM point to exclude
solutions in which multiple nuisance parameters---mostly for hadronic
corrections---simultaneously deviate strongly from prior expectations.

In total, there are 19 parameters $\alpha^i_j$ to describe $B\to K^*$ form
factors and most of them are strongly correlated. But it is well known that
strong correlation leads to poor sampling as it can cause the random-walk Markov
chains to spend an excessive amount of  time in regions of low probability and
thus produce spurious peaks. To mitigate this issue, we perform a fit to
form-factor constraints without any experimental data and use the resulting
covariance matrix to transform parameters such that the new parameters are
uncorrelated.

In all but three cases, the Markov chains then give reliable
results. But when we analyze scenarios with $\wilson{S,P} \ne 0$ and
all experimental constraints, strong correlations appear again. As a
final solution, we then use importance sampling with the initial
proposal function determined by a fit of a Gaussian mixture to the
MCMC samples within the variational Bayes
approximation~\cite{Jahn:2015}. As the posterior is unimodal and
closely resembles a Gaussian, only a few Gaussian components are
needed; i.e., 3 components proved optimal by the variational
approximation to the model evidence.

In the most challenging run with 62 parameters, we obtain a relative
effective sample size of only 0.038\%. We want to have enough
independent samples $N$ such that the 68\% region is determined with a
relative precision of about 1\%. As a rule of thumb, we consider the
``relative error of the error'' given by $1/\sqrt{2
  N}$~\cite[ch. 37]{Agashe:2014kda}.  In the 62D case, we compute a
total of $1.1\cdot10^6$ importance samples, update the proposal after
every $10^5$ samples, and combine all samples~\cite{Jahn:2015} such
that $N \approx 3500$ and the estimated relative error of the error is
1.2\% and thus good enough for our purposes.

To create the smooth marginal plots in Figures
\ref{fig:FHvsCT}--\ref{fig:scalar:SM-EFT}, we apply kernel density estimation
for both MCMC and importance samples using the fast \texttt{figtree}
library~\cite{morariu08figtree}. In the latter case, we additionally crop 500
outliers.

The prior or posterior predictive distribution of an observable $X$ within a
model $M$ in which $X=f(\vecth)$ is a definite function of the
parameters $\vecth$ {and} given as
\begin{align}
  \label{eq:predictive}
  P(X | M) &= \int \rmdx{\vecth} P(X|\vecth,M) P(\vecth|M)\nn
  &= \int \rmdx{\vecth} \delta(X - f(\vecth)) P(\vecth|M) \,.
\end{align}
We estimate $P(X|M)$ by computing $f(\vecth_i)$ for every sample $\vecth_i \sim
P(\vecth|M)$, then smooth as above.

%
%
%

%
%

\bibliographystyle{apsrev4-1}
\bibliography{references.bib}

\end{document}